\documentclass[11pt]{article}
\usepackage{indentfirst}
\usepackage{comment}
\usepackage{appendix}
\usepackage{graphicx}
\usepackage{amsmath}
\numberwithin{equation}{section}
\usepackage{amsfonts}
\usepackage{amssymb}
\usepackage{setspace}
\usepackage{booktabs}
\usepackage{cancel}
\usepackage{float}
\usepackage{sectsty}
\usepackage{bm,bbm}
\usepackage[T1]{fontenc}
\usepackage{textcomp}
\usepackage{listings}
\usepackage{caption}
\usepackage{subcaption}
\usepackage{stmaryrd} 
\usepackage{times}
\usepackage[backend=bibtex,sorting=none]{biblatex}
\usepackage{enumitem}
\usepackage{kotex}
\usepackage[font=footnotesize,format=plain,labelfont=bf]{caption}
\usepackage{xcolor}


\usepackage[colorlinks]{hyperref} 
\hypersetup{ 
    colorlinks=true,       
    linkcolor=red,          
    citecolor=blue,        
    filecolor=magenta,      
    urlcolor=cyan           
}

\newcommand{\eref}[1]{(\ref{#1})} 
\newcommand{\fref}[1]{Fig.~\ref{#1}}
\newcommand{\sref}[1]{Section~\ref{#1}}
\newcommand{\bF}{\mathbf{F}}

\begin{document}

\begin{center}
\textbf{\Large Comparative Analysis of Granular Material Flow: Discrete Element Method and Smoothed Particle Hydrodynamics Approaches}
\bigskip

\textbf{Jaekwang Kim$^{1}$, Hyo-Jin Kim$^{2}$, Hyung-Jun Park$^{3,*}$} \\
\bigskip
$^1$\textit{Department of Mechanical and Design Engineering, Hongik University, Sejong, South Korea}\\
$^2$Department of Korean Medical Science, College of Korean Medicine, \\Kyung Hee University, Seoul, Republic of Korea\\
$^3$\textit{School of Mechanical and Aerospace Engineering, Sunchon National University, Sunchon, South Korea}\\
hjpark89@scnu.ac.kr
\end{center}

\bigskip

\begin{center}
\textbf{Abstract }\\
\bigskip
\begin{minipage}{0.85\textwidth}
We compare two widely used Lagrangian approaches for modeling granular materials: the Discrete Element Method (DEM) and Smoothed Particle Hydrodynamics (SPH).
DEM models individual particle interactions, while SPH treats granular materials as a continuum using constitutive rheological models. 
In particular, we employ the Drucker–Prager viscoplastic model for SPH.
By examining key parameters unique to each method—such as the coefficient of restitution in DEM and the dilatancy angle in SPH—we assess their influence on two-dimensional soil collapse predictions against experimental results. While DEM requires computationally expensive parameter calibration, SPH benefits from a continuum-scale rheological model, allowing most parameters to be directly determined from laboratory measurements and requiring significantly fewer particles. However, despite its computational efficiency, viscoplastic SPH struggles to capture complex granular flow behaviors observed in DEM, particularly in rotating drum simulations. In contrast, DEM offers greater versatility, accommodating a broader range of flow patterns while maintaining a relatively simple model formulation.
These findings provide valuable insights into the strengths and limitations of each method, aiding the selection of appropriate modeling techniques for granular flow simulations.
\end{minipage}
\end{center}

\section{Background and introduction}

Granular materials such as sand, gravel, rice, and sugar, are everywhere in our daily lives.
The systems of granular materials show a variety of behaviors. They are strong enough to support medium. At the same time, 
when the granular medium is densely packed, it can flow apparently like a liquid. This is called dense granular flow. 
Such a granular medium is characterized by a collection of macroscopic particles of which size being typically greater than 100 $\mu$m.  
This size limitation restricts the types of interactions between particles; in a granular medium, the particles are non-Brownian and interact only through contact interactions such as friction and collisions. 
While research in modeling and predicting the mechanical behaviors of granular medium is motivated by its wide range of applications, from designing processing units like silos and rotating drums in industrial processes to predicting natural hazards such as landslides and rock avalanches in geophysics, 
the discrete nature of the particles introduces complexities into the flow that are not yet fully understood.

One of the most successful theoretical frameworks for modeling behavior of dense granular matters is the Discrete Element Method (DEM) developed by Cundall and Strack~\cite{Cundall}. 
Under the DEM framework, granular material is treated as an assembled system of distinct interacting particles, which have mass, velocity, position.
Then, contact-force models, which relate the pairwise interaction forces between paired particles to their geometric relationship, are used to track the positions of individual particles using Newton's second law. 
With rapidly increasing computational power, the DEM method has recently entered a new era, where a vast amount of research supports large-scale DEM simulations comparable to actual physical scales.
While some parameters of DEM contact models are grounded in macroscopic parameters (e.g., Young's modulus and Poisson's ratio), others still remain ambiguous, and reliable guidelines for calibrating them have yet to be clearly established~\cite{SPH_Bui:01,HLOSTA2018222}.

On the contrary, another perspective for granular flow considers the granular medium as a continuum with internal friction and plasticity.
In this approach, the counterpart of the contact models in DEM is the constitutive rheological models, which relate the stress within the continuum medium to the deformation and its rate.
Parameters of these constitutive models remain
macroscopic material properties, hence relatively easy to calibrate. However, the rheological model form often becomes extremely complicated, as the granular material can exist in different states under various mechanical environments.
Moreover, implementing such rheological models in numerical studies is a non-trivial task. 
This is because many flow scenarios of granular materials involve both large deformations and free surfaces,
which are detrimental to maintaining a high-quality grid in conventional numerical methods such as the finite element method. On the other hand, 
the Smoothed Particle Hydrodynamics (SPH)~\cite{Monaghan_2005_Review}, a gridless scheme initially developed for astrophysical problems, offers a promising technique to implementing the continuum perspective. Onwards, we will use the term SPH approach not only to refer to the numerical techniques but also to imply the continuum perspective of granular materials.

\begin{figure}
\begin{center}
\includegraphics[width=0.6\textwidth]{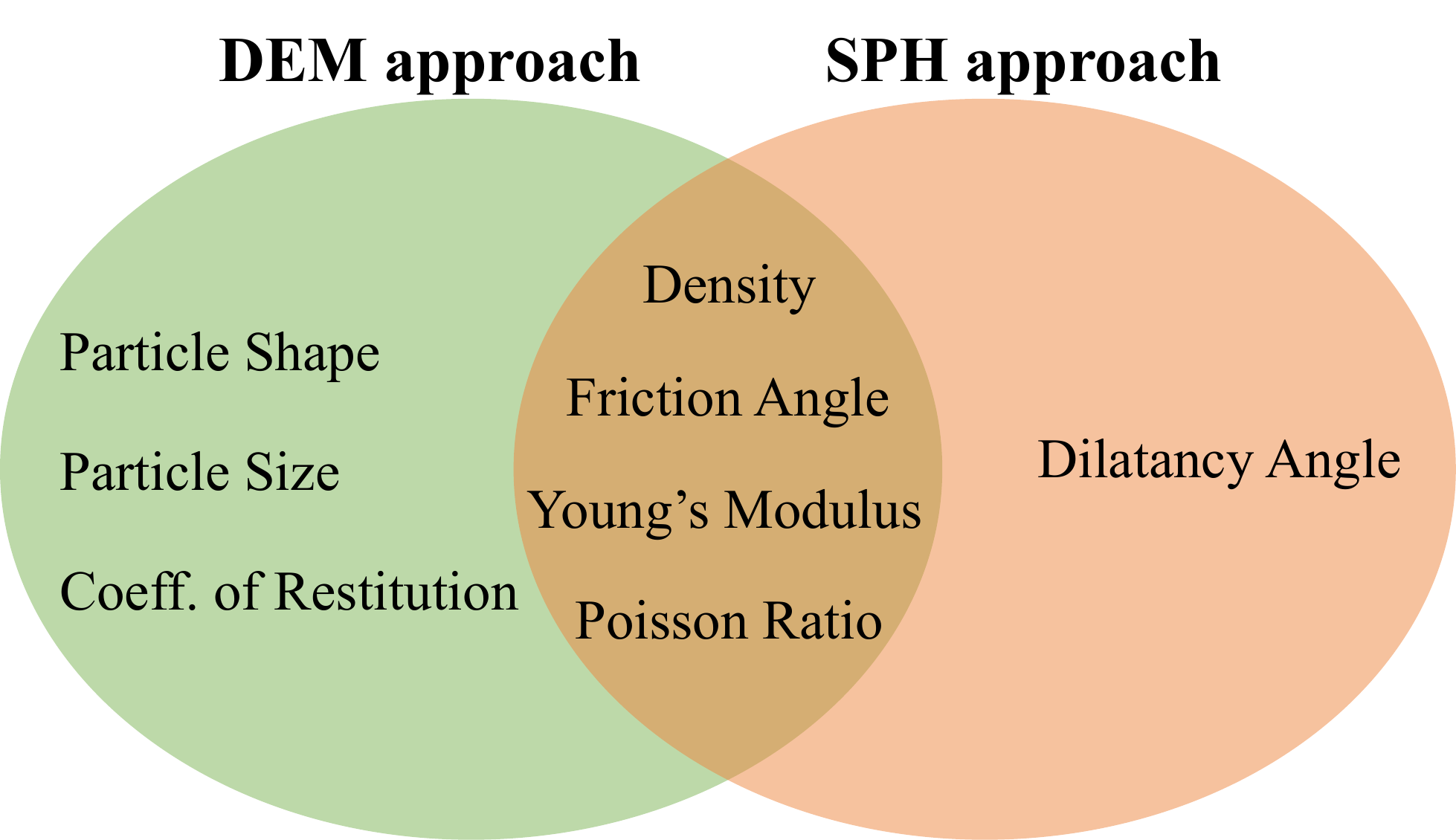}
\end{center}
\caption{
Parameters for the granular models in the DEM and SPH approaches. Note that some parameters are exclusively present in one approach but not in the other.
}\label{fig:schematic_sph_dem_compare}
\end{figure}

Of course, all models are approximate descriptions of underlying phenomena, and such model-form errors are always present in model-based predictions of granular flows. 
For example, conventional material properties handled in DEM and SPH approaches are summarized in \fref{fig:schematic_sph_dem_compare}. 
As illustrated, some parameters are exclusively present in one approach but not in the other. 
While the parameters of DEM contact models are based on the macroscopic properties of granular materials (e.g., Young's modulus, Poisson's ratio, etc.), they do not cover all properties. Typical contact models of the DEM approach do not include any parameter that can be directly related to the dilatant behavior of granular materials during shear deformation. 
In contrast, although such dilatant behavior can be readily incorporated in the continuum approach of SPH method, it does not account for all microscopic characteristics; the effects of the coefficient of restitution for two colliding granular particles as well as the effect of individual particle shapes and sizes are ignored in this method. 

In this regard, the goal of the present work is to investigate two different modeling approaches in particular flow scenarios of granular materials. 
Our aim is beyond a simple prediction comparison between two approaches, but to infer useful physical insight in each. 
To this end, we focus specifically on parameters that are exclusively present in each: the coefficient of restitution in the DEM approach and the dilatancy angle in the SPH method. 
In other words, this study investigates the exclusive role of these parameters in different flow scenarios and thus evaluate the capability of each approach.

Specifically, the first flow scenario involves large deformation and post-failure behaviors observed in the two-dimensional collapse of granular materials. Experimental observations in the literature~\cite{CollapseExperiment} of this benchmark flow will enable us to calibrate both DEM and SPH models to actual granular material behavior. Next, we extend our study to granular flows within a rotating drum.
The granular flow in a rotating drum is of great interest to researchers and finds widespread use across industries such as pharmaceuticals, mining, food processing, and chemical engineering. 
Typical applications include the mixing, coating, grinding, and drying of granular materials. 
The key difference in this flow scenario is the presence of a dynamic equilibrium at the free surface. Neither DEM nor SPH particles reach a true steady state; individual particles keep moving while the system maintains a stable free surface at a larger scale.

The rest of paper is structured as follow. 
\sref{sec:method} provides an overview of the two modeling approaches and details their implementation in the present numerical studies.  Section~\ref{sec:collapse} focuses on calibrating the parameters of the DEM contact model and the SPH rheological model using experimental data from two-dimensional collapse of rods in the literature. 
In Section~\ref{sec:drum}, the analysis is extended to granular flows in a rotating drum. Finally, Section~\ref{sec:conclusion} presents a discussion of the comparison studies and summarizes the main conclusions.

\section{Method}
\label{sec:method}

\subsection{The discrete element method for granular materials}

\begin{figure}
\begin{center}
\includegraphics[width=0.5\textwidth]{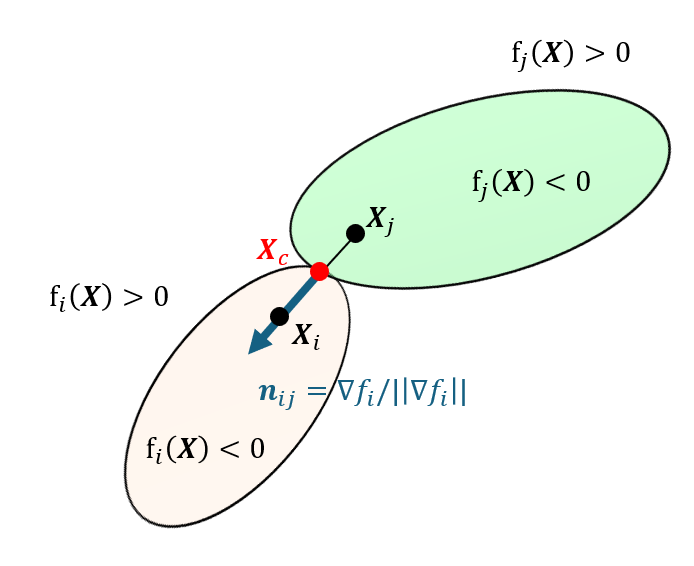}
\end{center}
\caption{Scheme of particle–particle contact for DEM approach. The contact point $\mathbf{X}_c$ and the overlap direction $\mathbf{n}_{ij}$ define the contact line. 
}\label{fig:DEM_contact_and_normal_overlap}
\end{figure}

DEM,introduced by Cundall and Strack~\cite{Cundall}, adopts the Lagrangian
approach in which the particles are treated as individual discrete entities.
The forces on the particles are computed using contact models at each time step through an explicit method, and the particles are tracked, with Newton's laws of motion determining their velocity and position. 
\begin{equation}
    m_i \ddot{\mathbf{x}}_i = \mathbf{F}^{\rm total}_i,
\end{equation}
where $m_i$ and $\mathbf{x}^{\rm total}_i$ are the mass and position of a particle center, and $\mathbf{F}_i$ is the total force acting on a particle $i$ resulting from particle-particle contact forces $\mathbf{F}^C_{ij}$ and external body forces $\mathbf{F}^{B}_i$ (e.g., gravity). 
The angular motions of particles are considered 
\begin{equation}
        I_i \dot{\mathbf{\omega}_i}= \mathbf{M^{\rm{total}}_i}
        \label{eqn:dem_angular_motion}
\end{equation}
where $I_i$ moment of inertia, $\omega$ angular velocity, and  $\mathbf{M}^{\rm total}_i$ is the total torque acting on the particle. 

The contact force $\mathbf{F}^C_{ij}$ between $i$-$j$ pair is composed of normal $\mathbf{F}^{n}_{ij}$ and tangential forces $\mathbf{F}^t_{ij}$ of the particles 
coming into contact with each other. 
\fref{fig:DEM_contact_and_normal_overlap} shows a schematic of particle-particle contact. The boundary of each particle is defined by a general shape function $\mathsf{f}_i(\mathbf{x}=(x,y,z))=0$. 
Note that the contact point $\mathbf{X}_c$ and the overlap direction $o_{n,ij}$ define the contact line. 

If the particles are non-spherical in shape, a contact detection algorithm should be formulated in terms of the optimization problem
\begin{equation}
\begin{aligned}
    &\mathrm{minimize} \quad \mathsf{f}_i(\mathbf{x})+\mathsf{f}_j(\mathbf{x}) \\
    &\mathrm{subject\; to} \quad \mathsf{f}_i(\mathbf{x})= \mathsf{f}_j(\mathbf{x}),
\end{aligned}
\label{eqn:dem_contact_detection}
\end{equation}
Applying the Lagrange multiplier approaches to  Eq. \eref{eqn:dem_contact_detection} results in a system of 4 equations with 4 unknowns~\cite{DEM_Cleary} that can be solved at each DEM time step for each pair of particles. 
For particles with a strongly asymmetrical shape, this requires  computationally expensive iterative methods (i.e. Newton’s method). 
The contact point $\mathbf{x}_c$ and the plane of overlap $o_{n,{ij}}$ are determined as the solution of the optimization problem.

\begin{table}
\begin{center}
\begin{tabular}{|c|c|c|}
\hline
 Symbol &  Parameter   & Equation \\
\hline
$R^*$ & Equivalent effective radius & $\frac{1}{R^*}=\frac{1}{R_i}+ \frac{1}{R_j}$ \\
 $m^*$ &Equivalent mass  & $\frac{1}{m^*}=\frac{1}{m_i}+ \frac{1}{m_j}$ \\
 $E^*$ &Equivalent Young's mass  & $\frac{1}{E^*}=\frac{1-\nu_i^2}{E_i}+ \frac{1-\nu_j^2}{E_j}$ \\
 $G^*$ &Equivalent shear modulus  & $\frac{1}{G^*}=\frac{2(2+\nu_i)(1-\nu_i)}{E_i}+ \frac{2(2+\nu_j)(1-\nu_j)}{E_j}$ \\
 $K_n$& Normal stiffness coefficient   & $K_n=\frac{4}{3}E^*\sqrt{R^* o_{n,ij}}$  \\
  $K_t$& Tangential stiffness coefficient  & $K_t=8 G^* \sqrt{R^* o_{n,ij}}$  \\
  $C_n$& Normal damping coefficient & 
  $2\sqrt{\frac{5}{6}}\frac{\ln (e)}{\sqrt{\ln^2(e)}+\pi^2} \sqrt{\frac{2}{3}K_n m^*} $  \\
   $C_t$& Tangential damping coefficient & 
  $2\sqrt{\frac{5}{6}}\frac{\ln (e)}{\sqrt{\ln^2(e)}+\pi^2} \sqrt{K_t m^*} $  \\
\hline
   $K_r$& Rolling stiffness coefficient  & $K_r=2.25 K_n \mu_r^2 R^{2}$ \\
   $C_r$ & The damping coefficient & 2$\eta_r\sqrt{I_r K_r}$ \\
   $\eta_r$ & The rolling viscous damping ratio & 0.3~\cite{EPSD}\\
$I_r$ & The equivalent moment of inertia &
    $I_r = \left[ 
    \frac{1}{I_i+m_i R_i^2}+ \frac{1}{I_j+m_j R^2_j}
    \right]^{-1}$
   \\
\hline
\end{tabular}
\caption{List of DEM microscopic material properties.
A set of fundamental material parameters is
$\vec{\Theta}^{DEM}=\{ \rho, E, \nu, \mu_s, \mu_r, e, c, R,l\}$.
The effective radius $R_i$ of a general shape material is determined by the reciprocal of the mean local curvature calculated at the nearest intersection point between the contact line and particle's surface. The rest parameters are derived quantities as a function of the normal overlap $o_{n,ij}$ of the contact plane.
}
\label{tab:dem_model_equations}
\end{center}
\end{table}

Next, we describe the contact friction force between two granular $ij-$particles, consisting of a set of homogeneous particles, of which $i$-particle has the Young's modulus $E^i$, Poisson ratio $\nu^i$. Also, $e$ denotes the coefficient of restitution between the particles.  
The contact force $\mathbf{F}^{C}_{ij}$ generated by particle collisions is determined based on the extent of normal overlap $o_{n,ij}$ and tangential overlap $o_{t,ij}$. 
In this work, Hertz's theory~\cite{DEM_Jonhson} is used to determine the normal force, while Mindlin's model~\cite{DEM_Sullivan} calculates the tangential force. 
This is written as 
\begin{equation}
\mathbf{F}^C_{ij}=\left\{
\begin{aligned}
& \bF^{n}_{ij}+\bF^{t}_{ij}, \quad \rm if \;  o_{n,ij} \geq 0\\
& 0, \quad \rm if \; o_{n,ij} < 0 ,
\end{aligned}
\right.
\end{equation}
with
\begin{equation*}
    \bF^{n}_{ij} = -K_n o_{n,ij}+ C_n \mathbf{v}^{rel}_n
\end{equation*}
and
\begin{equation}
    \bF^{t}_{ij} = \min\{ |-K_t o_{t,ij} + C_t \mathbf{v}^{rel}_n|, 
    \mu_s \mathbf{F}^{n}_{ij}\}. 
\end{equation}

In the above expressions, $K_n$ is the normal spring stiffness coefficient;
$C_n$ is the normal damping coefficient; 
$\mathbf{v}^{rel}_n$ is the normal relative velocities. 
The variables with the subscript $t$ denote tangential part of the corresponding quantities, while $\mu_s$ is the static coefficient of friction.

Furthermore, the equation of angular motion in Eq. \eref{eqn:dem_angular_motion}
is solved considering rolling resistances, 
which serves two primary roles: (i) dissipating energy during relative rotation in dynamic flow and (ii) providing packing support to maintain stability in the static phase. 
Using the EPSD model, the total torque $\mathbf{M}^{\rm total}$ of particle is written as 
\begin{equation}
\mathbf{M}^{\rm{total}}= \sum_{j} \mathbf{M}_{ij} + \mathbf{M}^r, 
\end{equation}
where $\bm{M}_{ij}= (\mathbf{x}_i - \mathbf{x}_j) \times \ \bF^{t}_{ij}$ is the torque acting on particle $i$ due to particle $j$, 
while $\bm{M}^r$ is the clipped torque due to rolling friction which is described as 
\begin{equation*}
    M^r = M^{k}+ M^{d}.
\end{equation*}
First, the spring torque $M^k$ is implemented in an incremental manner
\begin{equation}
    \Delta M^k = -K_r \Delta \theta_r, 
\end{equation}
where $\Delta \theta_r$ is the incremental relative rotation between particles and $K_r$ is the rolling stiffness.
The spring torque at time $t+\Delta t$ is found as
\begin{equation*}
        \ \min\{ |M^k_t|+ \Delta M^k, 
    \mu_r R^* F^n\}. 
\end{equation*}
On the other hand, the viscous damping torque $M^d$ is implemented as
\begin{equation}
M^d_{t+\Delta t}= -C_r \dot{\theta}_r,
\label{eqn:viscous_damping_torque_DEM}
\end{equation}
where $C_r$
is the damping coefficient.

Table~\ref{tab:dem_model_equations} presents the expressions for the DEM microscopic material properties in terms of the fundamental model parameter $\vec{\Theta}^{DEM}=\{ \rho,E, \nu, \mu_s, \mu_r, e, c, R,l\}$. 
It should be noted that $\mu_s$ and $\mu_r$ are microscopic parameters that need to be fitted to reproduce  experimental trends of granular flow, which has a bulk friction angle $\phi$ at the continuum scale. 
Additionally, some parameters, such as $e$, may require specialized experiments for accurate measurement~\cite{HLOSTA2018222}.

For time integration of DEM, the Verlet algorithm is commonly used~\cite{DEMtimeStpper}. 

\subsection{A continuum approach using the SPH}

In this section, we describe a continuum approach for granular flows, of which implementation relies on SPH.
We begin by introducing the former, and its numerical implementation follows. 
For illustration, throughout this section, we adopt Einstein notation, using Greek-letter superscripts to denote the components of vectors and tensors.
Here, we also remark the list of fundamental SPH model parameters $\vec{\Theta}^{SPH}=\{\rho, E,\nu,\phi, \psi, c\}$, where $\rho$ is the density, $\phi$ is the friction angle, $\psi$ is the dilatancy angle and $c$ is the cohesion. Other parameters which will be introduced below are derived quantities from  $\vec{\Theta}^{SPH}$.

\subsubsection{The constitutive model}
A densely packed granular medium can either flow or resist movement, depending on the magnitude of the external force. 
An appropriate class of constitutive models for such behavior is the viscoplastic models in plasticity theory. 
Below, we briefly summarize the theory of plasticity, explaining how the Cauchy stress tensor $\sigma$ of the material should be expressed in terms of strain rate. 
Following this, we present the specific expression of $\sigma$ for granular materials using the non-associated Drucker–Prager model. 
The main content of this section is adapted and reorganized from Ref.~\cite{SPH_Bui:01} to suit the purpose of the present paper.

The total strain rate tensor $\dot\epsilon$ for an elastic–perfectly plastic model is often divided into two parts: one part of elastic strain rate tensor $\dot\epsilon_{e}$
and the other part of plastic strain rate tensor $\dot\epsilon_{p}$
\begin{equation}
    \dot\epsilon = \dot\epsilon_e + \dot\epsilon_p. 
    \label{eqn:total_strain_plasticity}
\end{equation}

The generalized Hooke's law can be used to calculate the elastic strain rate tensor $\dot\epsilon_{e}$ 
\begin{equation}
    \dot\epsilon_e^{\alpha \beta} =\frac{\dot{s}^{\alpha \beta}}{2G}+
    \frac{1-2\nu}{3E}\dot{\sigma}^{\gamma \gamma}\delta^{\alpha \beta},
    \label{eqn:generalized_hooke_law}
\end{equation}
where $\dot{s^{\alpha\beta}}=\dot\sigma^{\alpha \beta} -(1/3)\dot{\sigma}^{\gamma\gamma}\delta^{\alpha\beta}$ is the deviatoric stress rate tensor and $\delta^{\alpha\beta}$ is Kronecker’s delta.
Note that in Eq. \eref{eqn:generalized_hooke_law} the superscripts $\alpha,\beta$ are free indexes that designate each component of tensors and $\gamma$ is a repeated index of Einstein notation, i.e. $\dot{\sigma}^{\gamma\gamma}= \dot\sigma^{xx}+\dot\sigma^{yy}+\dot\sigma^{zz}$. 

The plastic strain rate tensor is determined by applying the plastic flow rule
\begin{equation}
    \dot\epsilon^{\alpha\beta}_p = \dot{\lambda} \frac{\partial \mathcal{G}}{\partial \sigma^{\alpha \beta}}
    \label{eqn:plastic_flow_rule}
\end{equation}
where $\dot\lambda$ represents the rate of change of the plastic multiplier $\lambda$, which depends on the stress state and load history.
Moreover, $\mathcal{G}$ is the plastic potential, which is assumed to be perpendicular to the plastic strain increment in stress space. In other words, $\mathcal{G}$ characterizes the relationship between the stress state and plastic strain in the material.

It is assumed that yielding occurs when a certain combination of the stress components reaches a critical value, indicating the existence of a yield criterion $\mathcal{F}(\sigma)-k = 0$, where $k$ is some constant.
Defined in stress space,  
$\mathcal{F}$ limits the elastic regime of material
and thus called the yield surface.  
Then, the value of $\lambda$ can be calculated based on the fact  
that plastic deformation proceeds as long as the stress state remains
on the yield surface
\begin{equation}
    d\mathcal{F}=\frac{\partial \mathcal{F}}{\partial \sigma^{\alpha\beta}}d\sigma^{\alpha \beta}=0,
    \label{eqn:plasticity_consistency}
\end{equation}
which is known as the consistency condition. 
In general, the plastic multiplier $\lambda$
is zero during elastic loading ($\mathcal{F} \leq 0$) and 
plastic unloading ($d\mathcal{F}<0$), 
while $\lambda >0$ during plastic deformation ($\mathcal{F}=0$ and $d\mathcal{F}=0$). 
 
Substituting Eqs. \eref{eqn:generalized_hooke_law} and \eref{eqn:plastic_flow_rule}
to Eq. \eref{eqn:total_strain_plasticity}, 
following the procedures in Appendix~\ref{sec_app:Cauchy_Stress}, 
we obtain the following explicit form of $\sigma^{\alpha \beta}$
\begin{equation}
    \dot{\sigma}^{\alpha \beta}=2G{{\dot{e}}^{\alpha \beta }}+K{{\dot{\epsilon }}^{\gamma \gamma }}{{\delta }^{\alpha
\beta }}-\dot{\lambda }\left[ \left( K-\frac{2}{3}G \right)\frac{\partial \mathcal{G}}{\partial
{{\sigma }^{mn}}}{{\delta }^{mn}}{{\delta }^{\alpha \beta }}+2G\frac{\partial \mathcal{G}}{\partial
{{\sigma }^{\alpha \beta }}} \right], 
\label{eqn:cauchy_stress_results}
\end{equation}
where 
\begin{equation}
    K=\frac{E}{3(1-2\nu)}
    \label{eqn:plasticity_K}
\end{equation}
and
\begin{equation}
    G=\frac{E}{2(1+\nu)}
    \label{eqn:plasticity_G}.
\end{equation}
In Eq. \eref{eqn:cauchy_stress_results}, $\dot{e}^{\alpha \beta}$ denotes the deviatoric strain rate tensor $\dot{e}^{\alpha \beta}= \dot\epsilon^{\alpha \beta}- \dot\epsilon^{\gamma\gamma}\delta^{ \alpha \beta}/3$

Once the yield function $\mathcal{F}$ and plastic potential function $\mathcal{G}$
are defined for a specific material, and the total strain rate tensor $\dot\epsilon$
is provided, using the consistency condition in Eq. \eref{eqn:plasticity_consistency}, $\dot\lambda$ can be determined
\begin{equation}
    \dot\lambda = \frac{2G\dot \epsilon^{pq}\frac{\partial \mathcal{F}}{\partial \sigma^{pq}}+\left( K-\frac{2G}{3}\right)\dot{\epsilon}^{\gamma\gamma}\frac{\partial \mathcal{F} }{\partial \sigma^{pq}} \delta^{pq}}{2G\frac{\partial \mathcal{F}}{\partial \sigma^{pq}}\frac{\partial \mathcal{G}}{\partial \sigma^{pq}}
    +\left( K-\frac{2G}{3} \right)
    \frac{\partial \mathcal{F}}{\partial \sigma^{mn}}\delta^{mn} \frac{\partial \mathcal{G}}{\partial \sigma^{pq} } \delta^{pq}
    }, 
    \label{eqn:lambda_flowrate_general}
\end{equation}
where $m,n,p$ and $q$ are dummy indexes. 

In this work, we employ the non-associated Drucker–Prager model that 
$\mathcal{F}$ as 
\begin{equation}
    \mathcal{F}^{DP}(I_1, J_2) =\sqrt{J_2}+\alpha_{\phi}I_1 - k_c =0,
    \label{eqn:DP_yield_crietria}
\end{equation}
where $I_1$ is the first invariant of $\sigma$ 
\begin{equation*}
    I_1 = \sigma^{xx} + \sigma^{yy} + \sigma^{zz},
\end{equation*}
and $J_2$ is the second invariant of the deviatoric part of $s$
\begin{equation*}
    J_2=\frac{1}{2}s^{\alpha\beta}s^{\alpha \beta},
\end{equation*}
while $\alpha_{\phi}$ and $k_c$ are material constants determined by the friction angle $\phi$,
\begin{equation*}
    \alpha_\phi = \frac{\tan\phi }{\sqrt{9+12\tan^2 \phi}},
\end{equation*}
\begin{equation*}
    k_c = \frac{3c}{\sqrt{9+12\tan^2\phi}}.
\end{equation*}
Secondly, we consider the following non-associated plastic potential $\mathcal{G}^N$
\begin{equation}
    \mathcal{G}^{N}(I_1, J_2)=\sqrt{J_2}+3I_1 \sin\psi,
    \label{eqn:DP_nonassociated_rule}
\end{equation}
where $\psi$ is the dilatancy angle of granular materials. 

Substituting Eqs. \eref{eqn:DP_yield_crietria} and \eref{eqn:DP_nonassociated_rule} into Eq. \eref{eqn:cauchy_stress_results}, and 
introducing the Jaumann stress rate 
\begin{equation}
\hat{\sigma}^{\alpha \beta}= \dot{\sigma}^{\alpha\beta} -\sigma^{\alpha \gamma} \dot{\omega}^{\beta \gamma} - \sigma^{\gamma \beta}\dot{\omega}^{\alpha \gamma}
\label{eqn:Jaumann_stress}
\end{equation}
in the place of $\sigma$,
the final form the stress-strain relationship for an elastic–perfectly plastic material is derived as follow
\begin{equation}
    \dot{\sigma}^{\alpha\beta} -\sigma^{\alpha \gamma} \dot{\omega}^{\beta \gamma} - \sigma^{\gamma \beta}\dot{\omega}^{\alpha \gamma}= 2G \dot{e}^{\alpha\beta} + K \dot{\epsilon}^{\gamma \gamma}\delta^{\alpha \beta} - \dot\lambda \left[ 
    3 K \alpha_{\psi} \delta^{\alpha \beta} + \frac{G}{\sqrt{J_2}}s^{\alpha \beta}
    \right],
    \label{eqn:sigma_evolution}
\end{equation}
in which $\alpha_{\psi}$ is material constants determined by the dilatancy angle angle $\psi$.

In Eqs. \eref{eqn:Jaumann_stress} and \eref{eqn:sigma_evolution}, $\dot\omega^{\alpha \beta}$ is spin rate tensor
\begin{equation*}
    \dot{\omega}^{\alpha \beta} = \frac{1}{2}\left(
\frac{\partial v^{\alpha}}{\partial x^\beta}-
\frac{\partial v^{\beta}}{\partial x^\alpha}
    \right)
\end{equation*}
Note that the Jaumann stress rate in Eq. \eref{eqn:Jaumann_stress} was introduced to construct a stress rate that is invariant with respect to rigid-body rotation upon a large deformation. 
Similar approaches are also widely applied in constitutive models of viscoelastic materials~\cite{JKim:2023,Renardy:2021}.
Moreover, substituting Eqs. \eref{eqn:DP_yield_crietria} and \eref{eqn:DP_nonassociated_rule} into Eq. \eref{eqn:lambda_flowrate_general},
the rate of change of plastic multiplier $\dot\lambda$ is written as 
\begin{equation}
    \dot\lambda = \frac{3 \alpha_{\phi} K \dot{\epsilon}^{\gamma\gamma} + (G/\sqrt{J_2}) s^{\alpha\beta} \dot{\epsilon}^{\alpha\beta}}{27 \alpha_{\phi}  K \sin \psi +G }. 
\label{eqn:lambda_evolution}
\end{equation}

\subsubsection{The SPH scheme}

The application of viscoplastic models, particularly in scenarios involving large deformations and post-failure, poses significant challenges to grid-based methods such as the finite element method~\cite{JKim:2018,JKim:2019} and the finite volume method~\cite{kim2023application}. 
In contrast, the SPH method, one of the longest-established meshless numerical techniques, has proven efficient for implementing viscoplastic models. 
Due to its Lagrangian and adaptive nature, SPH handles large deformations and post-failure behavior more effectively than grid-based methods. 
In the following, we briefly outline the SPH methodology used in this work, which has been implemented with an in-house code and tested in earlier works for various applications, 
including heat transfer~\cite{Park:2024} and free surface flows~\cite{kim2023direct, Park:2023}.

In SPH, the partial differential equations for the continuum mechanics are first converted into equations of motion of particles, which carry field variables (such as mass, density, stress tensor, etc) and their motions are updated with the material velocity. 
For instance, the SPH framework uses Lagrangian formulation of mass and momentum conservation: 
\begin{equation}
    \frac{D\rho}{Dt} = -\frac{1}{\rho}\frac{\partial v^\alpha}{\partial x^\alpha}
    \label{eqn:SPH_rhoupdate}
\end{equation}
\begin{equation}
    \frac{Dv^{\alpha}}{Dt} = \frac{1}{\rho}\frac{\partial \sigma^{\alpha\beta}}{\partial x^\beta} + g^{\alpha}
    \label{eqn:SPH_vupdate}
\end{equation}
where $g^\alpha$ of Eq. \eref{eqn:SPH_vupdate} is the component of gravitational acceleration.  
Note that the plastic flow of granular materials can be explicitly simulated using the evolution equations of Cauchy stress $\sigma$ of Eq. \eref{eqn:sigma_evolution} and the plastic multiplier $\lambda$ of Eq. \eref{eqn:lambda_evolution}.

\begin{figure}
\begin{center}
\includegraphics[width=1\textwidth]{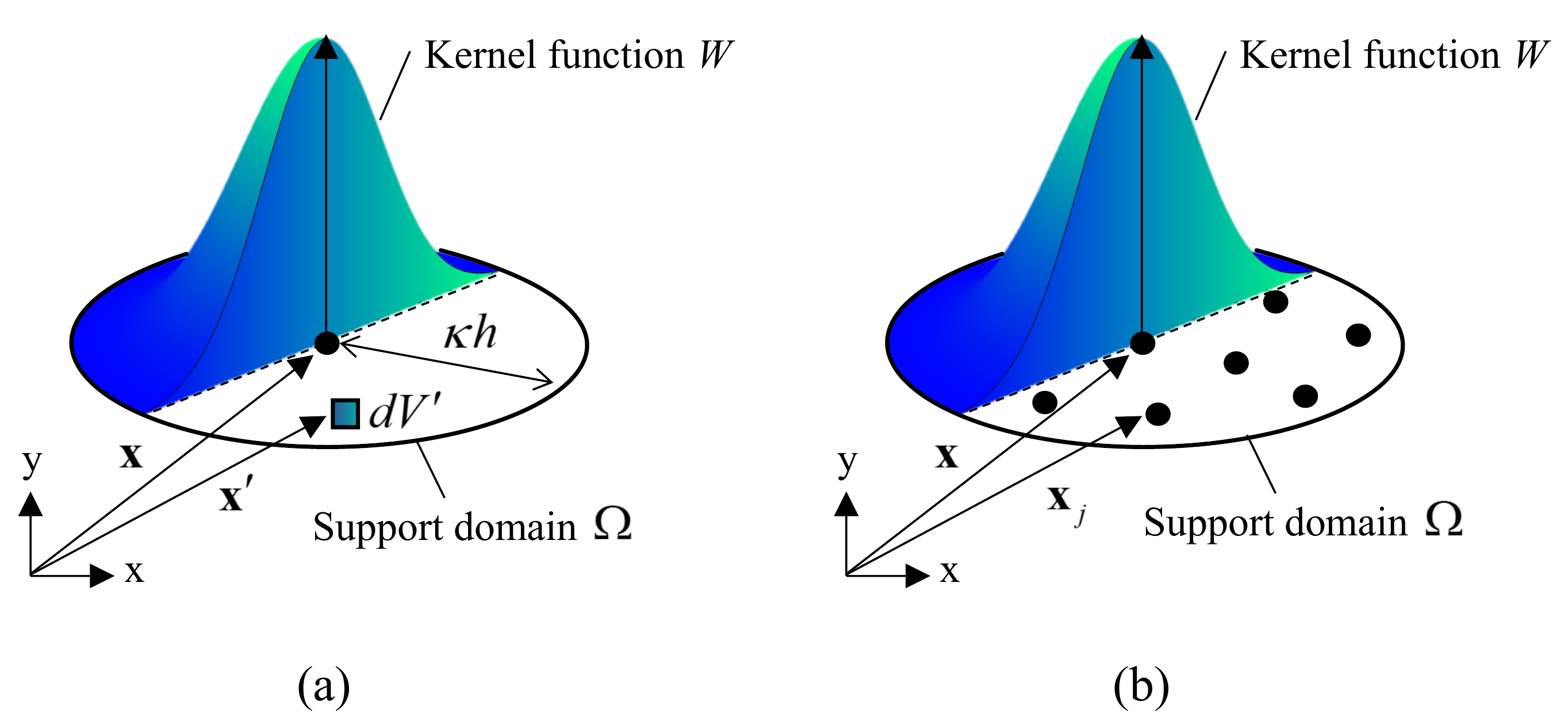}
\end{center}
\caption{Schematic configuration for SPH approximation: (a) kernel approximation and (b) particle approximation.
}\label{fig:SPH_discretization}
\end{figure}

The key idea of SPH framework during this process is to approximate a field function $f({\bf{x}})$ defined on domain $\Omega$ using a finite number $N$ of interpolating particles. 
These particles carry material properties (such as velocity, density, and stress and move with the material velocity), 
which are calculated through the use of an interpolation process over its neighboring
particles.
The interpolation is based on the integral representation of $f$ and its gradient:
\begin{equation}
    f({\bf{x}})\, \cong \,\int_\Omega  {f({\bf{x'}})\,W({\bf{x}} - {\bf{x'}},h)\,dV'},
    \label{eqn:fx_kernel_approxi}
\end{equation}
\begin{equation}
    \nabla f({\bf{x}})\, \cong \, - \int_\Omega  {f({\bf{x'}})\,\nabla W({\bf{x}} - {\bf{x'}},h)\,dV'},
    \label{eqn:del_fx_kernel_approxi}
\end{equation}
where $dV'$ represents the differential volume, 
$W$ a kernel function, and $h$ the smoothing length that defines the influence domain of $W$.  
As depicted in \fref{fig:SPH_discretization}, 
the radius of the support domain for $W$ is determined by $\kappa h$, where $\kappa $ is a parameter dependent on the choice of kernel function. In this work, we use the Wendland kernel function 
\begin{equation}
W_{\mathrm{wend}}(q)= 
\left\{
\begin{aligned}
    &(1-\frac{q}{2})^4(1+2q), \quad  0 \leq q \leq 2\\
    &0, \quad 2<q
\end{aligned}
\right.
\label{eqn:SPH_wendlend_kernel}
\end{equation}
resulting in $\kappa =2$~\cite{huIncompressibleMultiphaseSPH2007}.
For a choice of even fucntion for $W$, the integral representations are of second-order accuracy~\cite{liu2003smoothed}. 
Next, the continuous integral representations, i.e. Eqs. ~\eref{eqn:fx_kernel_approxi} and \eref{eqn:del_fx_kernel_approxi} are discretized as a summation over particles in the support domain 
\begin{equation}
    f({{\bf{x}}_i})\, \approx \sum\limits_{j = 1}^N {f({{\bf{x}}_j})\,W_{ij}\frac{{{m_j}}}{{{\rho _j}}}},
    \label{eqn:fx_particle_approxi}
\end{equation}
\begin{equation}
    \nabla f({{\bf{x}}_i})\, \approx \sum\limits_{j = 1}^N {f({{\bf{x}}_j})\,[\nabla _i W_{ij}] \frac{{{m_j}}}{{{\rho _j}}}},
    \label{eqn:del_fx_particle_approxi}
\end{equation}
with 
\begin{equation}
    W_{ij}=W({\bf{x}}_i -{\bf{x}}_j,h)\quad {\rm{and}} \quad \nabla_i W_{ij}= \frac{\partial W_{ij}}{\partial \mathbf{x}_i} = \left( \frac{\mathbf{x}_i -\mathbf{x}_j}{r}
    \right)\frac{\partial W_{ij}}{\partial r}. 
    \label{eqn:partical_kernel}
\end{equation}
In Eq.~\eref{eqn:partical_kernel}, $r=|\mathbf{x}_i - \mathbf{x}_j|$ denote the relative distance between particles. From Eq.~\eref{eqn:fx_kernel_approxi} to Eq.~\eref{eqn:partical_kernel}, the gradient operation results vary depending on the type of field function. Therefore, the more general Gibbs notation is used for convenience.

For example, using Eq. \eref{eqn:del_fx_particle_approxi} and gradient of unity, the velocity gradient, which is an ingredient for evaluating strain rate $\dot{\epsilon}$ and spin rate $\dot{\omega}$, can be calculated using the symmetrized form 
\begin{equation}
    \frac{\partial v^\alpha_i}{\partial x^\beta} = \sum\limits_{j = 1}^N \frac{{{m_j}}}{{{\rho _j}}}(v^\alpha_j -v^{\alpha}_i)\cdot \frac{\partial W_{ij}}{\partial x^{\beta}_i},
\end{equation}
to ensure that the gradients of a constant velocity ﬁeld vanish.

As used in recent SPH literature~\cite{bui2021smoothed}, 
the conservations of mass in Eq. \eref{eqn:SPH_rhoupdate} and momentum in Eq. \eref{eqn:SPH_vupdate} are evaluated using discretized forms 
\begin{equation}
    \frac{D \rho_i}{D t} = \sum\limits_{j = 1}^N m_j (v^\alpha_i -v^{\alpha}_j)\cdot \frac{\partial W_{ij}}{\partial x^{\alpha}_i},
\end{equation}
and 
\begin{equation}
    \frac{Dv^{\alpha}_i}{Dt} = 
    \sum\limits_{j = 1}^N m_j 
    \left(
\frac{\sigma^{\alpha\beta}_i}{\rho_i^2}
+
\frac{\sigma^{\alpha\beta}_j}{\rho_i^2}
    \right)
   \frac{\partial W_{ij}}{\partial x^{\beta}}+ b^{\alpha}. 
\label{eqn:SPH_vupdate_discretized}
\end{equation}
Eq. \eref{eqn:SPH_vupdate_discretized} arises from the chain-rule of derivatives
\begin{equation*}
    \frac{1}{\rho_i}\frac{\partial \sigma^{\alpha\beta}_i}{\partial x^\beta}= \frac{\partial }{\partial x^\beta}\left( \frac{\sigma^{\alpha\beta}_i}{\rho_i}
    \right)
    + \frac{\sigma^{\alpha\beta}_i}{\rho^2_i}\frac{\partial \rho_i}{\partial x^\beta}, 
\end{equation*}
which is devised to effectively evaluate the divergence of stress. 
Discretizations of soil constitutive equation in Eq. \eref{eqn:sigma_evolution} in the SPH framework are also straightforward
\begin{equation}
\frac{D\sigma^{\alpha\beta}_i}{Dt}= \sigma^{\alpha \gamma}_i \dot{\omega}^{\beta \gamma}_i + \sigma^{\gamma \beta}_i\dot{\omega}^{\alpha \gamma}_i+ 2G \dot{e}^{\alpha\beta}_i + K \dot{\epsilon}^{\gamma \gamma}_i\delta^{\alpha \beta}_i - \dot\lambda_i \left[ 
    3 K \alpha_{\phi} \delta^{\alpha \beta}_i + \frac{G}{\sqrt{J_2}}s^{\alpha \beta}_i
    \right],
\end{equation}
and 
\begin{equation}
    \frac{D\lambda_i}{Dt} = \frac{3 \alpha_{\phi} K \dot{\epsilon}^{\gamma\gamma}_i + (G/\sqrt{J_2}) s^{\alpha\beta}_i \dot{\epsilon}^{\alpha\beta}_i}{9 \alpha_{\phi}  K +G }. 
\end{equation}

While the momentum conservation in Eq. \eref{eqn:SPH_vupdate_discretized} must be stabilized to avoid numerical oscillation and tensile instabilities in actual implmentation of SPH scheme~\cite{feng2021large, nguyen2017new}, we leave more details on such numerical techniques in
Appendix~(\ref{sec_app:SPH_stabilizaton}).

For time integration, predictor-corrector schemes~\cite{NumericalMethod} are used.

\section{Collapse of granular columns}
\label{sec:collapse}

The first benchmark flow is large deformation and post-failure of non-cohesive soil in a rectangular channel, which is also referred to as dam break phenomena. 
The problem domain is illustrated in the schematic shown in \fref{fig:Dam_collapse_domain}.
The phenomenon has been extensively studied to gain a deeper understanding of dense granular flow.

\begin{figure}
\begin{center}
\includegraphics[width=0.8\textwidth]{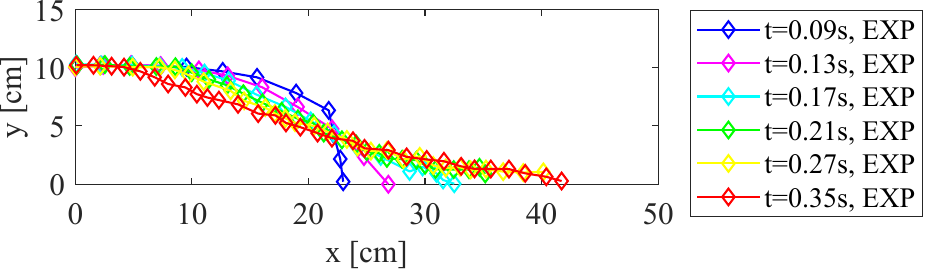}
\end{center}
\caption{The collapse free surface of the granular column as observed in experiments of Ref.~\cite{CollapseExperiment}. The initial shape has a height of 100 mm and a width of 200 mm.}
\label{fig:Collapse_EXP_freesurface}
\end{figure}

One important goal of this test is to calibrate parameters of numerical models. In particular, we adopted data available in Ref.~\cite{CollapseExperiment}. 
In their work, the shape of 2D granular column collapses is investigated using 50 mm long aluminum rods with two diameters: 1.6 mm and 3.0 mm, mixed at a weight ratio of 3:2. Note that the large aspect ratios of the rods, 31.25 and 16.67 (length to diameter), ensure the validity of the two-dimensional assumption. 
The properties of 2D soil model considered in the experiment~\cite{CollapseExperiment} are summarized in Table~\ref{tab:properties_of_experiment}.
Here, the density $\rho$ of granular bulk is measured using the total weight divided by initial volume.
\fref{fig:Collapse_EXP_freesurface}, generated using digitized data points from photographs taken during the experiment, demonstrates the transient free surface during the collapse. 
In the figure, the $x$-direction is defined as the runout direction and the $y$-direction as the height of the initial rectangle. 

\begin{table}
\begin{center}
\begin{tabular}{c|c|c|c} 
\hline
Name    & Values reported in the experiment  & In DEM & In SPH \\
\hline
Density $\rho$  & 20.4  $[\rm{kN/m^3}]$ & O & O\\
Young's modulus $E$  & 5.84  [MPa] &O & O\\
Poisson's ratio $\nu$ & 0.3 & O & O\\
Bulk Friction Angle $\phi$ & 21.9$^{\circ}$ & X & O\\ 
Dilation Angle $\psi$&  5-7$^{\circ}$ & X &O \\
The static coefficient of friction $\mu_s$ & Not Available & O & X \\
The rolling coefficient of friction $\mu_r$ &  Not Available & O &  X \\
The coefficient or restitution $e$ &  Not Available & O &  X \\
Cohesion $c$  & 0 [kPa] & O & O \\
Size $R$  & 1.6 to 3.0 [mm] & O& X\\
Rod length $l$  & 50 [mm] & O & X \\
\hline
\end{tabular}
\caption{
Material properties for the quasi-2D soil model used in the granular column collapse experiment by Nguyen et al.~\cite{CollapseExperiment}. The third and fourth columns indicate whether each parameter is implemented in the respective numerical schemes. }
\label{tab:properties_of_experiment}
\end{center}
\end{table}

\subsection{The DEM simulation}

The DEM approach for modeling the evolution of free surface is implemented using an open-source software \texttt{LIGGGHTS}~\cite{Liggght}.
In particular, to simulate plane strain (i.e., quasi-2D) of the experiment, the DEM simulations were conducted in 3D but with large aspect ratio, 12:1. 
To this end, we employed superquadric particles in the DEM simulation. 
The general shape equation of a superquadric object, 
with the coordinate origin defined at particle center, 
is given as 
\begin{equation*}
    \mathsf{f}(x,y,z) = \left(
    \left| \frac{x}{a}
    \right|^{n_2}
    +
    \left| \frac{y}{b}
    \right|^{n_2}
    \right)^{n1/n2}+
    \left| \frac{z}{c}
    \right|^{n_1}-1=0,
\end{equation*}
where $a, b$, and $c$ are the half-length of the particles along its principal axes, and $n_1$ and $n_2$ are blockiness parameters. 
The superquadric shape is understood as 
an extension of spheres and ellipsoids~\cite{Barr}. 
A cylinder is obtained if $n_1=2$ and $n_2 \gg 2$. 
In general, it is reported that the numerical stability of DEM simulation using superquadratic particles decreases with increase of blockiness parameters $n_1$ and $n_2$~\cite{DEM_superquadratic}. 
In our work, we used $n_1=2$ and $n_2=20$. 
 
The list of DEM parameters are $\vec{\Theta}^{DEM}=\{ \rho,E, \nu, \mu_s, \mu_r, e, c, R,l\}$. Among them 
$\vec{\Phi}^{DEM}=\{\mu_s,\mu_r, e\}$ are highly uncertain, 
while rest of parameters $\vec{\Theta}^{DEM}-\vec{\Phi}^{DEM}=\{\rho, E, \nu, c, R,l \}$ can be selected based on the experimentally reported values, as shown in Table~\ref{tab:properties_of_experiment}, assuming the system consists of homogeneous particles.
Additionally, it should be noted that the DEM model formulation does not allow for the straightforward incorporation of the material's dilatancy behavior, as there exists no DEM parameter that explicitly corresponds to the dilatancy angle $\psi$.

\begin{figure}
\begin{center}
\includegraphics[width=0.8\textwidth]{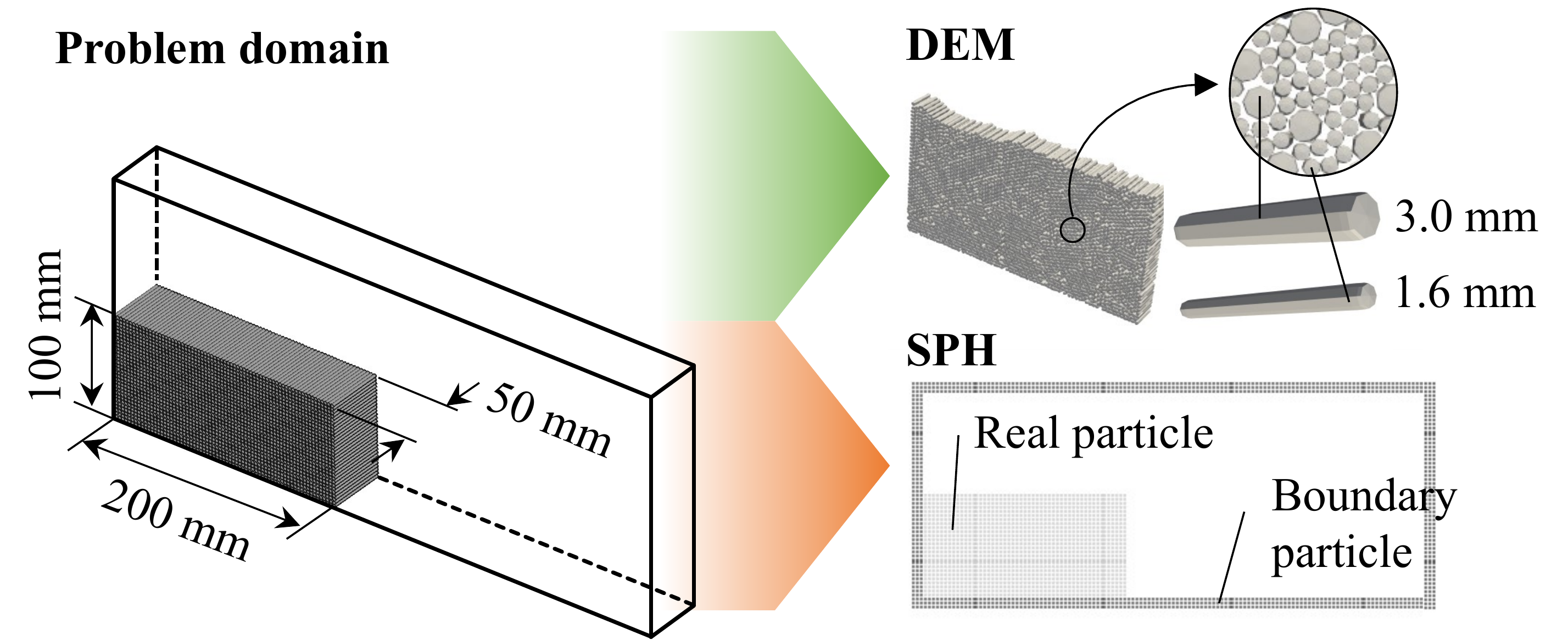}
\end{center}
\caption{Domain description for granular column collapse problem.}
\label{fig:Dam_collapse_domain}
\end{figure}

\begin{figure}
\begin{center}
\includegraphics[width=0.8\textwidth]{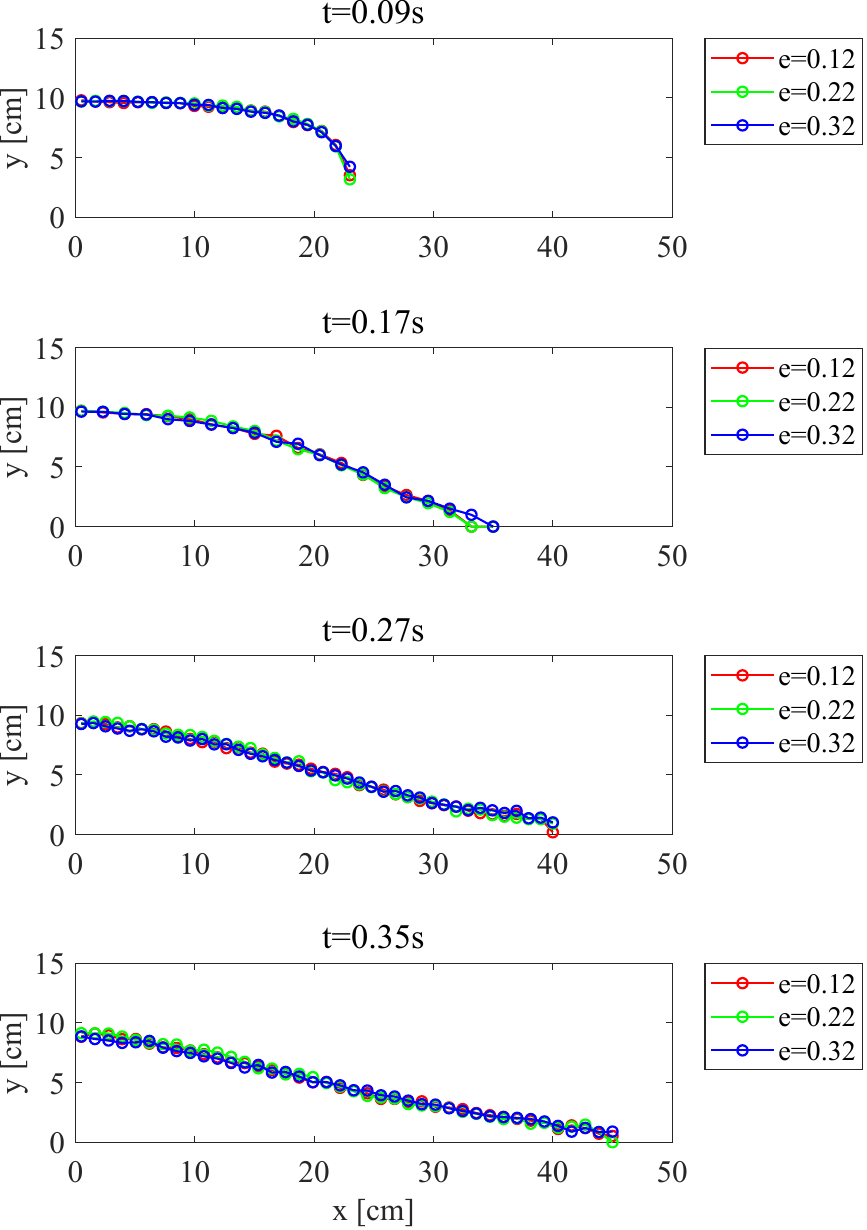}
\end{center}
\caption{Evolution of free surface during collapse for different values of $e$. In the simulation $\{\mu_s,\mu_r\}$ is fixed as $\{0.4, 0.0\}$. The results show that both transient and final free surface shapes are not significantly affected by the value of $e$.}
\label{fig:Collapse_DEMtransient_with_e}
\end{figure}

The initial configuration of densely packed granular system in the DEM simulation, prior to collapse, is prepared using the following procedures.
First, five solid wall boundaries are constructed in the shape of cuboids, leaving the top surface open. 
Then, a total of $N$ initial rod particles, comprising a mixture of rods with diameters of 1.6 mm and 3.0 mm, are aligned 15 cm above the cuboid and mixed at a weight ratio of 3:2 to match the experimental conditions. 
A small artificial gravity, $g^*=0.1$ $\rm m/s^2$, is applied to all particles to settle them into the cuboid. 
Once the particles fill the cuboid, an artificial wall, with its normal in the $y$-direction, is used to compress the system, creating a densely packed granular configuration.
Finally, the system is allowed to relax over 0.1 s. The total particle number, $N=6,300$, was iteratively determined to match the initial height of $h=100$ mm in the experiment after relaxation. 
The initial configuration is shown in \fref{fig:Dam_collapse_domain}. 
Finally, the system is subjected to the true gravitational acceleration $g=9.81$ $\rm m/s^2$ and the side wall located at $x=20$ cm is suddenly removed, triggering the collapse of the densely packed system.
Throughout the DEM simulation, time step size is fixed $\Delta t = 1.25\times 10^{-5}$ s, after verifying that the results remain consistent for smaller values of $\Delta t$. 
Using 16 cores, the DEM simulation takes 12 hours to complete, including the relaxation phase before the collapse begins.

\begin{figure}
\begin{center}
\includegraphics[width=0.8\textwidth]{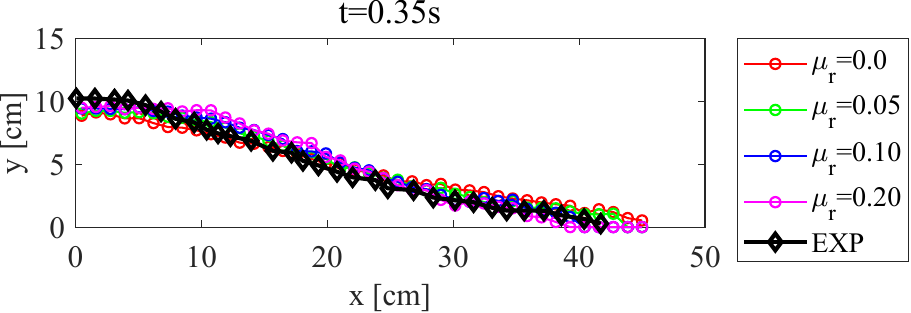}
\end{center}
\caption{Comparison of the final free surface between the experiment and the DEM prediction for different values of $\mu_r$.
As $\mu_r$ increases, less material flows outward.}
\label{fig:Collapse_DEMfinal_with_mu_r}
\end{figure}

The calibration steps for the DEM parameter is known to be a challenging and non-unique process~\cite{ANGUS2020290}. 
In this work, we calibrated the unknown DEM parameters $\vec{\Phi}^{DEM}$ using the following procedures.
First, we initialize $\vec{\Phi}^{DEM}=\{\mu_s,\mu_r,e\}=\{0.4, 0.0, 0.12\}$. Then, we investigate the influence of $e$ by increasing its value. 
\fref{fig:Collapse_DEMtransient_with_e} demonstrates the transient free surface of granular collapse for different values of $e$.
As noted reported in Ref.~\cite{Chau:1999}, it is observed that both the final and transient surface shapes exhibit little sensitivity to this parameter. In other words, the collapse flow is not useful for calibrating $e$ of the DEM parameters. 
Therefore, we fix $e=0.12$ for the following analyses
and focus on $\mu_s$ and $\mu_r$. 
While a general rule of thumb for choosing these parameters is to reproduce experimental observations corresponding to the measured bulk friction angle $\phi$, multiple combinations of $\{\mu_s, \mu_r\}$ have been reported to achieve this~\cite{ANGUS2020290}.
This lead us to fix $\mu_s =0.4$ and iteratively adjust 
$\mu_r$ to find the optimal combination that best matches 
the final shape of the experimental surface.
\fref{fig:Collapse_DEMfinal_with_mu_r} compares the evolution of the free surface between the experiment and the DEM prediction with $\mu_r$ while $e$ and $\mu_s$ are fixed as 0.12 and 0.4, respectively.
We observed that as $\mu_r$ increases, more volume accumulates near the fixed wall, while less material flows outward.

\begin{figure}
\begin{center}
\includegraphics[width=0.8\textwidth]{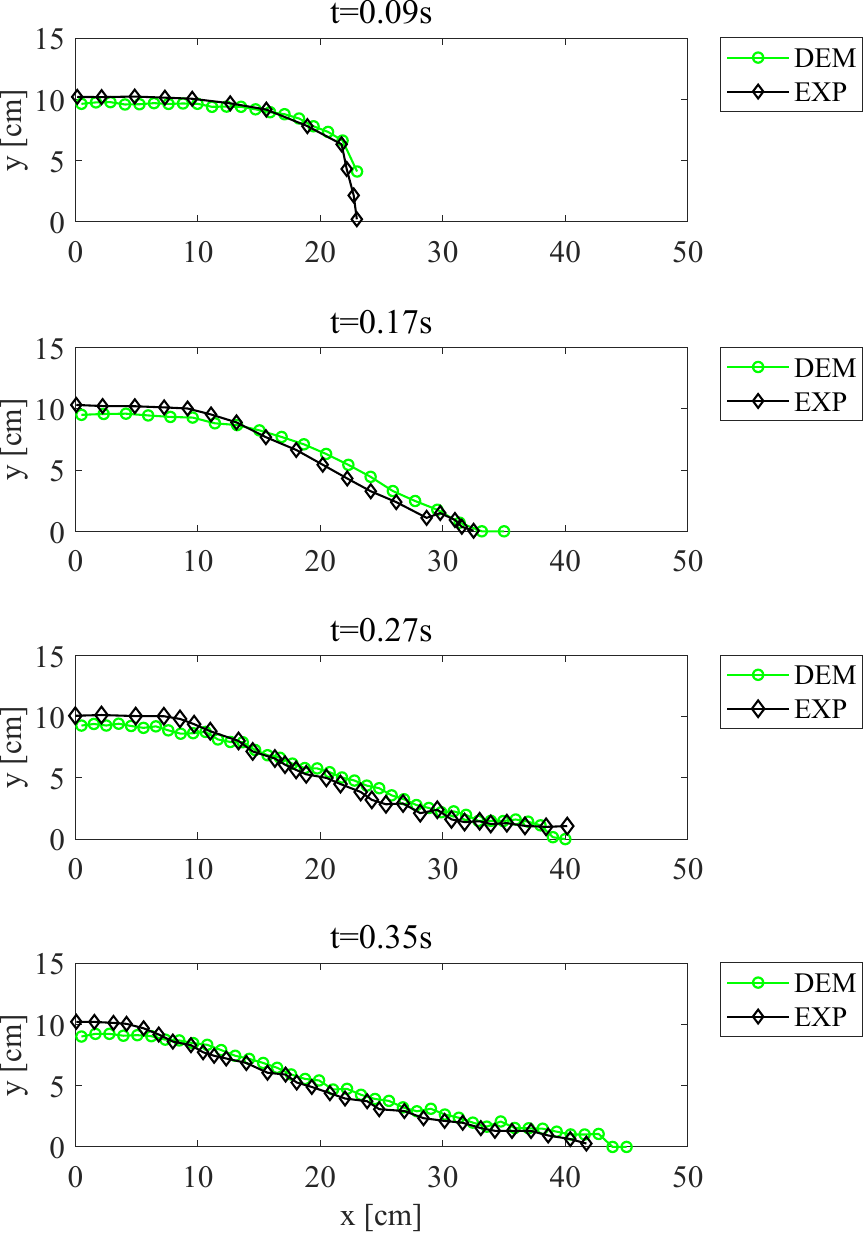}
\end{center}
\caption{Free surface comparison between experimental observation and the DEM prediction with calibrated parameters $\vec{\Phi}^{DEM}=\{\mu_s,\mu_r, e\}=\{0.4, 0.05,0.12\}$ during collapse. The overall transient shapes show good agreement.}
\label{fig:Collapse_DEMandEXP}
\end{figure}

To quantitatively assess the degree of free surface agreement, we consider the following integral:
\begin{equation}
    \int^\infty_{0} |\mathcal{S}^{EXP}(x)- \mathcal{S}^{DEM}(x)| dx, 
    \label{eqn:freesurface_comparison}
\end{equation}
where $\mathcal{S}^{EXP}$ and $\mathcal{S}^{DEM}$ denote the free surfaces of the final collapsed shape observed in the experiment and DEM simulation, respectively. 
Numerical interpolation and integration are used evaluate Eq. ~\eref{eqn:freesurface_comparison} and the values are summarized in 
Table~\ref{tab:dem_parameter_calibration}. 
With this, we determined $\vec{\Phi}^{DEM}=\{\mu_s,\mu_r,e\}=\{0.4, 0.05, 0.12\}$,
and the final transient free surface plots are compared with the experiment in \fref{fig:Collapse_DEMandEXP}.
We conclude that the overall transient shapes show good agreement after the parameter calibration. 

\begin{table}
\begin{center}
\begin{tabular}{c|c}
\hline
$\mu_r$  [-]  & Discrepancy~\eref{eqn:freesurface_comparison} [cm$^2$] \\
\hline
0.0  & 28.17\\
0.05 & 20.60 \\
0.1 & 22.49 \\
0.2 & 26.10 \\
\hline
\end{tabular}
\vspace*{5mm}
\caption{Discrepancy in the free surface profile between the experiment $\mathcal{S}^{EXP}$ and the DEM simulation $\mathcal{S}^{DEM}$. The metric used is Eq.~\eref{eqn:freesurface_comparison}. The parameters $e$ and
$\mu_s$ are fixed at 0.12 and 0.4, respectively, while $\mu_r$=0.05 shows the best match. }
\label{tab:dem_parameter_calibration}
\end{center}
\end{table}

\subsection{The SPH simulation}

\begin{figure}
\begin{center}
\includegraphics[width=0.8\textwidth]{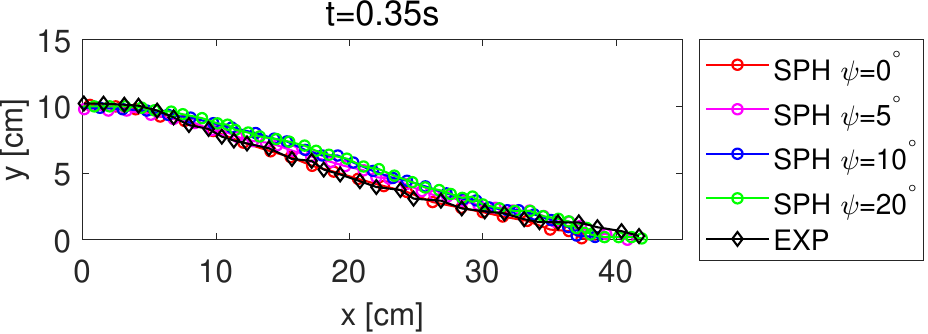}
\end{center}
\caption{The final free surface for different values of the dilatancy angle $\psi$ in SPH simulations.
The final volume increases with an increase in $\psi$.}
\label{fig:CollapseSPHDilatancyAngle}
\end{figure}

\begin{figure}
\begin{center}
\includegraphics[width=0.8\textwidth]{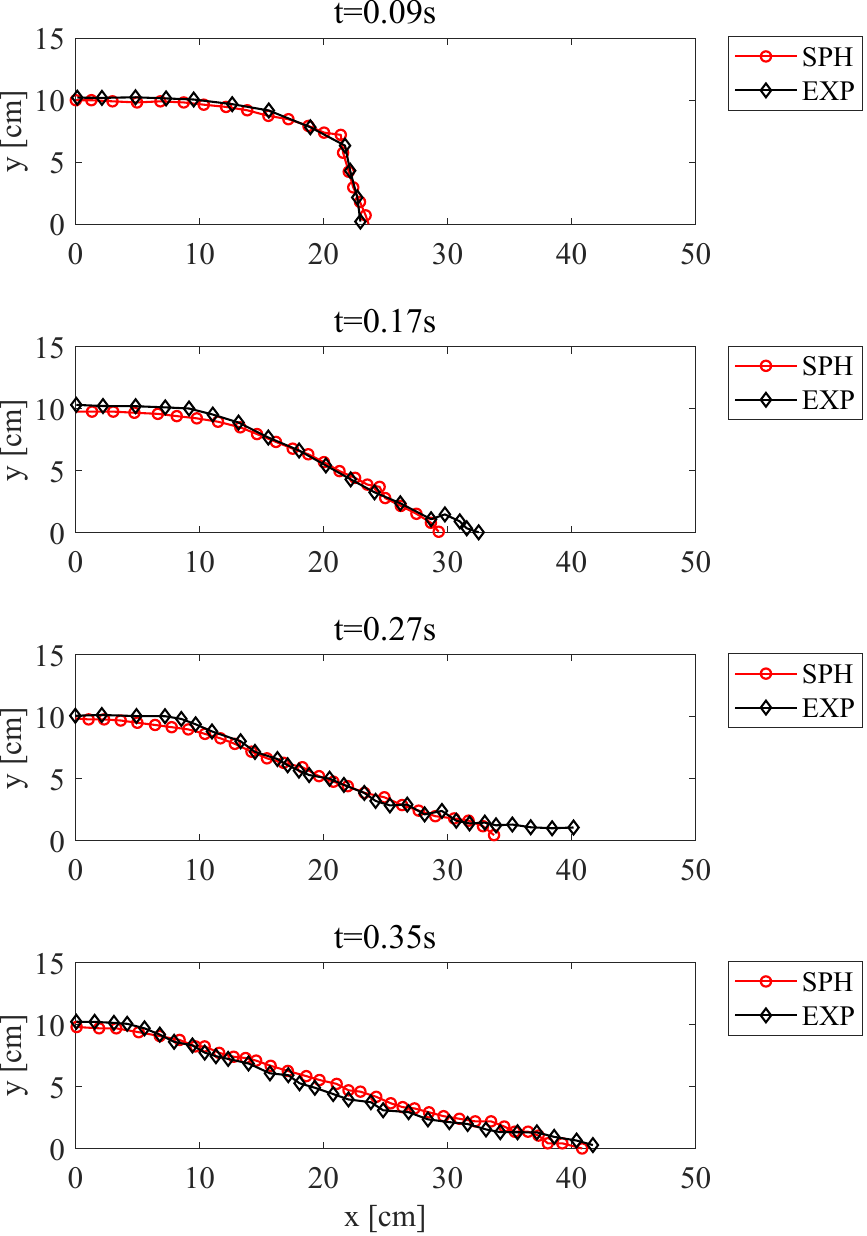}
\end{center}
\caption{Free surface comparison between experimental observation and the SPH prediction with $\psi=5^{\circ}$ during collapse. While the overall transient shapes show good agreement, the SPH predicts less lateral flow and a longer runout distance.}
\label{fig:CollapseSPHandEXP}
\end{figure}

Next, we revisit the same problem using the viscoplastic constitutive model at the continuum scale and the SPH numerical scheme.
Assuming a perfectly plane strain condition, the two-dimensional SPH scheme is implemented using an in-house code.
Recall that the large aspect ratio in the experiments (at least 16.67) enables a direct comparison with the two-dimensional SPH scheme.

Based on classical plasticity theory, the SPH model assumes continuous material properties, and, therefore, most of SPH model parameters $\vec{\Theta}^{SPH}=\{\rho, E, \nu, \phi, \psi, c\}$ can be explicitly determined from the material properties reported in the experiment. The only exception is the particle size $R$, which is not accounted for in the SPH constitutive model. 

In a similar vein, intricate packing procedures, as required in the DEM configuration, 
are unnecessary for SPH. 
Since SPH particles are not physical entities but merely represent the properties of the surrounding medium, simply distributing a sufficient number of particles in this work—in a structured manner is sufficient.
In our case, we set $N=1,326$, resulting in an approximately 4.75-to-1 mapping between the DEM particles and SPH interpolating particles.
Using only a single-core, the SPH simulation completes in just 220 minutes. This represents a significant reduction in computation time compared to the DEM simulation.

In the SPH test, we begin by examining the role of the dilatancy angle $\psi$. 
Recall that the parameter is reported as $\psi=5$ to $7^{\circ}$ in the experiment, whereas
it was not possible to incorporate it into the DEM simulation. 
The final free surface shapes for different values of the dilatancy angle $\psi$ in SPH simulations are shown in \fref{fig:CollapseSPHDilatancyAngle}. 
From the plot, it is observed that the final volume increases with an increase in $\psi$,  which can be attributed to the more pronounced dilatant behavior of granular materials during shear.
Such dilatant behavior is more quantitatively confirmed using the volume per unit depth of the final shapes, calculated as the area underneath the curves in \fref{fig:CollapseSPHDilatancyAngle} and summarized in Table~\ref{tab:sph_psi_effect}.
The discrepancy in the final free surface compared to the experiment is also reported in the same table.
While the results suggest that the SPH simulation best matches the experimental data when $\psi=0^{\circ}$, \fref{fig:CollapseSPHDilatancyAngle} shows that this case significantly underestimates the run-out distance ($\approx$ 38 cm), whereas the experimental value is nearly $\approx$ 42 cm. 
In contrasts, the case $\psi=5^{\circ}$ accurately predicts the runnout distance, and the overall discrepancy 18.4 $\rm cm^2$ in the free surface is smaller than in all previous DEM simulations ($>20$ $\rm cm^2$). 
Thus, we conclude to fix $\psi = 5^{\circ}$, as reported in Table~\ref{tab:properties_of_experiment}.

In \fref{fig:CollapseSPHandEXP}, we compare the transient evolution of the free surface predicted by the SPH scheme with experimental observations. At all times, the two results show stronger agreement compared to the previously presented DEM prediction. This suggests that the material's dilatancy is a significant factor in accurate free surface predictions, whereas particle size is not a crucial parameter.

\begin{table}
\begin{center}
\begin{tabular}{c|c|c }
\hline
$\psi$ [$^{\circ}$]    & Volume per unit depth [cm$^2$] & Discrepancy [cm$^2$] \\
\hline
0 & 193.93 & 8.74\\
5 & 214.61 & 18.4 \\
10 & 224.78 & 27.0 \\
20 & 230.13 & 27.9 \\
\hline
\end{tabular}
\vspace*{5mm}
\caption{Volume per unit depth (area underneath the curves) of the final shape shown in  \fref{fig:CollapseSPHDilatancyAngle}.
}
\label{tab:sph_psi_effect}
\end{center}
\end{table}

\begin{figure}
\begin{center}
\includegraphics[width=0.8\textwidth]{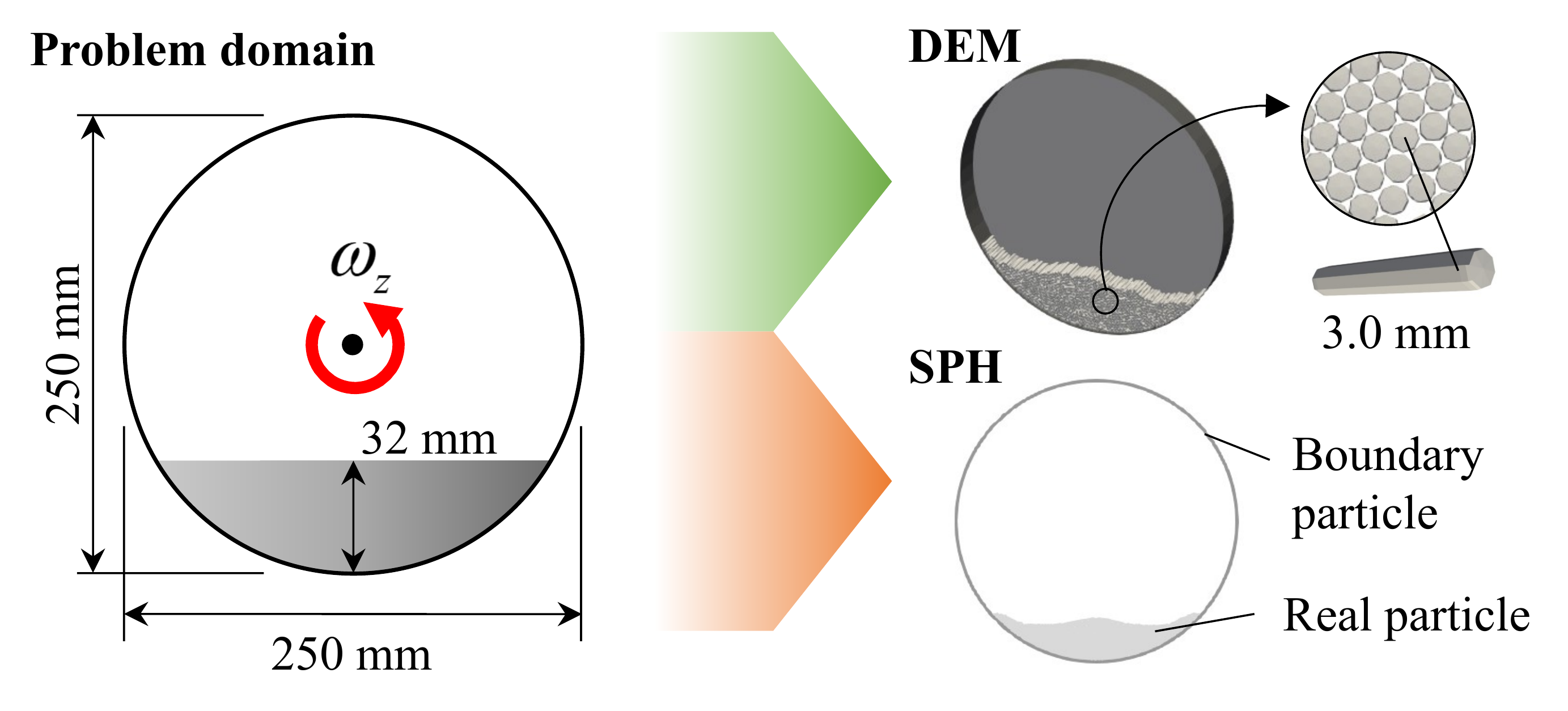}
\end{center}
\caption{Domain description for rotary drum problem.}
\label{fig:Rotary_drum_domain}
\end{figure}

\section{Another case: granular flows in rotating drum}
\label{sec:drum}

\begin{figure}
\begin{center}
\includegraphics[width=0.8\textwidth]{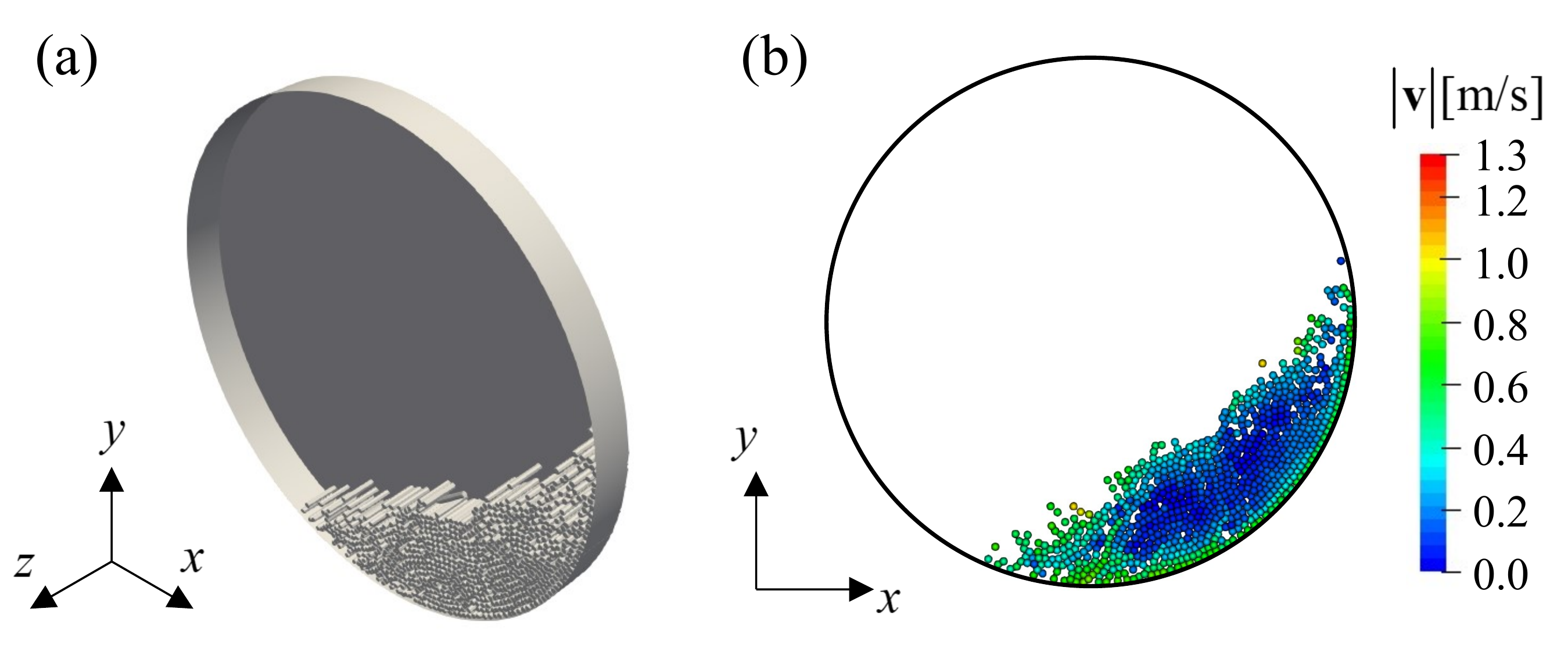}
\end{center}
\caption{DEM simulation snapshots for a rotary drum with an angular velocity $w_z=2\pi $ rad/s and the coefficient of restitution $e=0.12$. (a) A 3D representation of the superquadric element used in this study. (b) Velocity distribution projected onto the $z=0$ plane.  }
\label{fig:DEM_Rotary_drum_snapshot}
\end{figure}

\begin{figure}
\begin{center}
\includegraphics[width=0.8\textwidth]{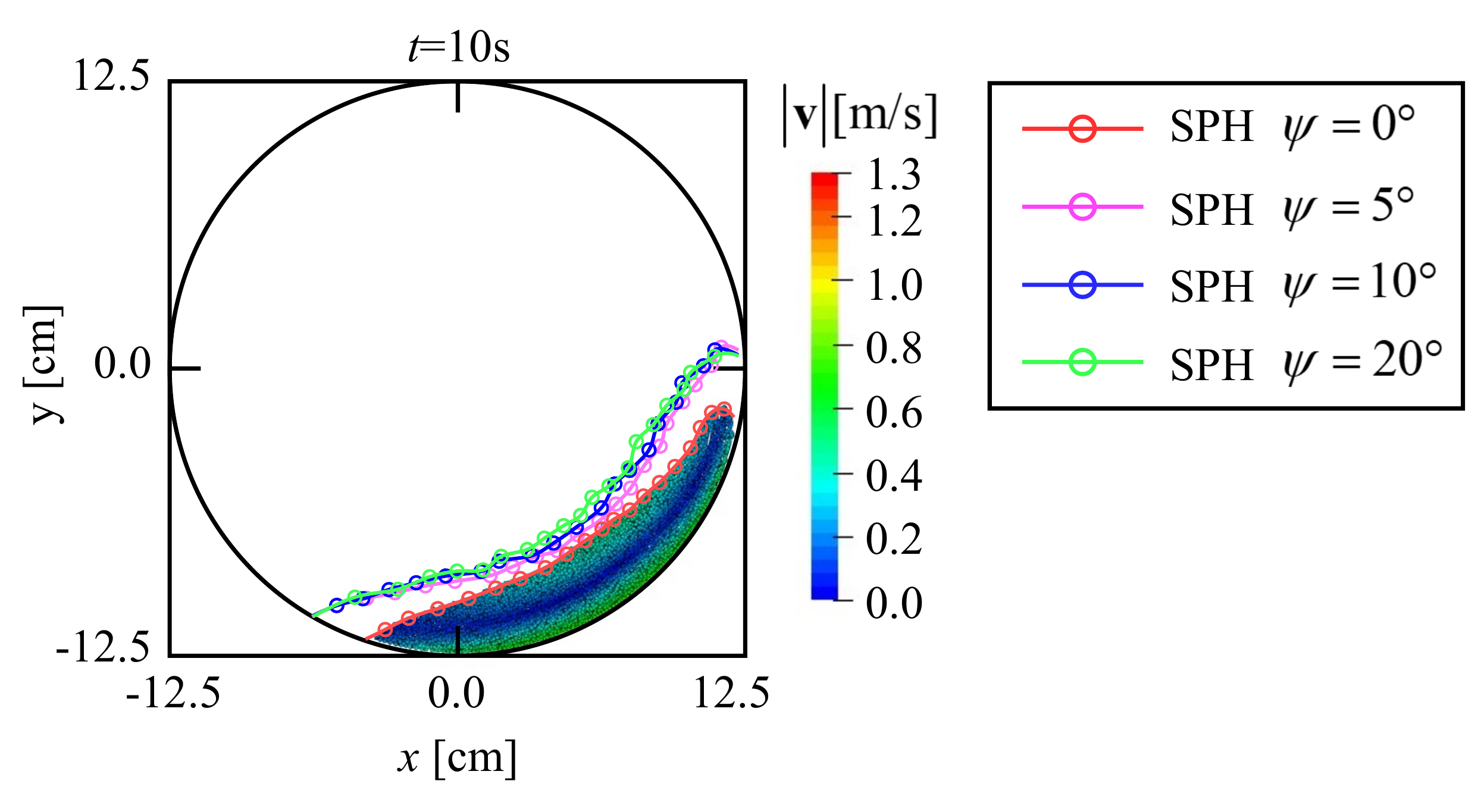}
\end{center}
\caption{SPH simulation snapshots for rotary drum with angular velocity $w_z=2\pi $ rad/s  and comparison of the free surface according to the dilatancy angle $\psi$. The velocity distribution is plotted for the case $\psi=0^{\circ}$.}
\label{fig:SPH_rotary_drum_snapshots}
\end{figure}

\begin{figure}
\begin{center}
\includegraphics[width=0.8\textwidth]{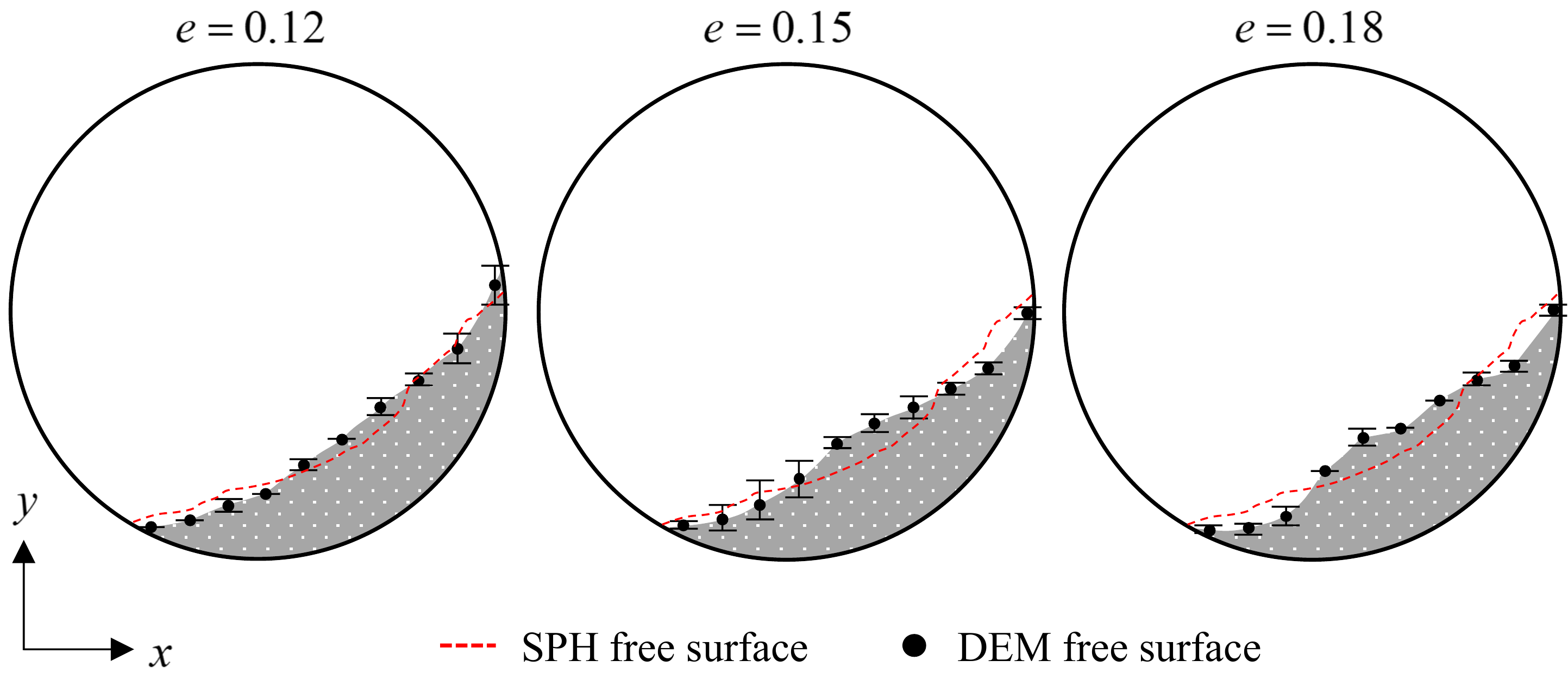}
\end{center}
\caption{Distribution of DEM particles in the drum rotating counterclockwise at $w_z=2\pi $ rad/s for different values of the coefficient of restitution $e$. The red dashed line represents the SPH free surfaces at $\psi=20^{\circ}$, plotted as a reference.}
\label{fig:DEM_Rotary_drum_freeSurf}
\end{figure}

As the second example, we analyze granular flow in a rotating drum. 
The flow scenario is illustrated with a schematic overview of the problem provided in \fref{fig:Rotary_drum_domain}.
In this study, we retain the quasi-two-dimensional setting.  
The flow configuration is as follow: 
A drum, which has a radius of 0.125 m, is partially filled with granular particles and rotates with a fixed angular velocity  $\omega_z =2\pi \rm $ rad/s around its central axis. 
As the drum rotates, the particles in contact with the inner wall experience frictional forces that cause them to move with the drum. However, once the gravitational force exceeds the frictional resistance, the particles start to detach and slide down along the free surface. 
Consequently, the shape profile is determined by the two competing effects of these forces.
The competing effects are often characterized by the Froude number~\cite{POURANDI2024120176,Juarez_2011} 
\begin{equation*}
\mathrm{Fr}=\sqrt{\frac{\omega_z^2 D}{2g}}
\end{equation*}
where $D$ is the drum diameter.
It is reported that different flow regimes manifest according to Fr ~\cite{RotatingDrum0}. 
According to the choice of $D$ and $\omega_z$, the Fr is estimated as 0.71. 

In this numerical experiment, we fixed the set of model parameters $\vec{\Theta}^{DEM}$ and $\vec{\Theta}^{SPH}$
as in the previous collapse test, since, unfortunately, we do not have experimental results for directly calibrating the DEM parameters.
Instead, we focus on investigating the respective role of coefficient of restitution $e$ of DEM and dilatancy angle $\psi$ of SPH, which distinguish the different capabilities of two approaches for the granular flow in a rotating drum.

First, the DEM simulation is set up as follows: similar to the previous case, a total of 3,200 rod particles are generated around the center of the static drum and allowed to fall onto the drum's wall boundaries.
This setup fills approximately 1/5 of the drum's volume.
After the system reaches a relaxed state, the drum wall begins to rotate at a constant angular velocity of  $\omega_z =2\pi \rm $ rad/s. 
It is observed that the system achieves a dynamic equilibrium at the free surface within $t=3.5$ s.
A snapshot of the DEM simulation at $t=3.7$ s is shown in 
\fref{fig:DEM_Rotary_drum_snapshot}, along with the velocity distribution  projected onto the $z=0$ plane. On the other hand, while initial SPH configuration is prepared in a similar manner, it only requires $N=2,601$ SPH particles to fill a comparable volume in the drum.

The snapshot of the SPH scheme and the velocity distribution for the case $\psi=0^{\circ}$ is shown in \fref{fig:SPH_rotary_drum_snapshots}.
The velocity distributions of both schemes exhibit similar trends:
maximum velocities are observed near the free surface or close to the drum wall, while the inner region shows minimal velocities. 
This pattern aligns with findings from studies on granular flow in rotating drums, where higher velocities are typically found at the surface layer due to active particle movement, 
and lower velocities prevail in the core region.

\begin{table}
\begin{center}
\begin{tabular}{c|c}
\hline
$\psi$ [$^{\circ}$]    & Volume per unit depth [mm$^2$] \\
\hline
0  & 4044.80\\
5 & 6405.90 \\
10 & 7078.40 \\
20 & 7343.47 \\
\hline
\end{tabular}
\vspace*{5mm}
\caption{For the rotary drum problem, volume per unit depth at equilibrium states for different values of $\psi$.}
\label{tab:sph_psi_effect_rotary}
\end{center}
\end{table}

In \fref{fig:SPH_rotary_drum_snapshots}, the influence of the dilatancy angle $\psi$ in the SPH scheme on the volume per unit depth at equilibrium states is also compared. A more quantitative analysis is summarized in  Table~\ref{tab:sph_psi_effect_rotary}. 
As observed in the previous case, the equilibrium volume increases with $\psi$, as the material expands more under shear. 
On the other hand, 
\fref{fig:DEM_Rotary_drum_freeSurf} compares the distribution of the free surface in the DEM simulation at equilibrium states for different values of $e$. 
Here, because characterizing a dynamically maintained surface is more challenging than a static one, the free surface of the DEM is plotted with uncertainties. The uncertainties are estimated using the standard deviation of the free surface location at three different time points after the system reaches equilibrium.

Overall, the free surface profile aligns best with the SPH results when $e$ is at its minimum value of 0.12.
As $e$ increases, the profile deviates differently compared to the effect of $\psi$ in the SPH scheme;
DEM particles begin to accumulate in certain areas, resulting in an uneven free surface.
We explain such trends with respect to $e$ at both micro and macro scales as follows: 
First, from a microscopic perspective, as $e$ increases, more kinetic energy from the drum wall is transferred to the upper layers through frictional forces, as illustrated in the following process. 
When the drum wall rotates, granular particles attached to the wall begin to move with it, gaining kinetic energy through frictional interactions. As these particles move along the wall, they collide with stationary particles in the upper layer, transferring kinetic energy upward. 
Consequently, at larger $e$, the influence of wall friction forces on the overall shape becomes more pronounced.
This contrasts with the downward gravitational body force, which acts uniformly on all particles without requiring collisions. While gravity tends to flatten the free surface, its effect remains independent of
$e$. 
From a macroscopic view, as $e$ increases, the elasticity of the granular medium as a bulk increases. As a result, a more energy can be stored in the granular medium, contributing to the accumulation of granular particles and leading to an uneven free surface profile.
Such elastic effects cannot be captured by the viscoplastic rheological models of the SPH scheme.

In conclusion, the numerical experiments suggest that the dilatancy angle $\psi$ of the SPH scheme and the coefficient of restitution $e$ of the DEM scheme plays unique role within its own framework, which cannot be reproduced by the other scheme.

\section{Conclusion}
\label{sec:conclusion}

In this work, we compared two common Lagrangian approaches for modeling granular materials: the Discrete Element Method (DEM), 
which models individual particle interactions, 
and Smoothed Particle Hydrodynamics (SPH), 
which treats the material as a continuum using plasticity theory.

By examining key parameters unique to each method
—such as the coefficient of restitution in DEM and the dilatancy angle in SPH— we first examined their impact on two-dimensional soil collapse predictions against experimental results. 
Since the model parameters of DEM are not macroscopic quantities typically used to characterize experimental results, they must be re-calibrated to fit observations. 
However, DEM simulations require a large number of particles, matching those in the actual system. This leads to significant computational costs. Consequently, model calibration becomes a time-consuming task.
On the other hand, SPH schemes employ continuum-scale rheological models, allowing most model parameters to be directly determined from laboratory-measured quantities. 
Additionally, because SPH requires far fewer particles than DEM, it may offer significant computational advantages.

However, these differences do not imply that one method is universally superior to the other. 
As demonstrated in granular flow simulations of a rotating drum, the SPH scheme within the viscoplastic regime fails to capture the diverse flow patterns observed in DEM simulations.
Of course, there exists more sophisticated rheological model in continuum scales~\cite{JKim:2023,JKim:2019,Jkim2021} which can be potentially implemented using the SPH scheme.
Whether elastoviscoplastic (EVSP) models can reproduce such patterns remains an open question. However, their implementation is challenging due to the complexity of their rheological structure.
In contrast, DEM can reproduce various flow patterns without sacrificing its relatively simple model formulation, as it provides more adjustable parameters to fit a wide range of scenarios.
Nevertheless, our SPH scheme demonstrated at least same accuracy in reproducing the experimental observations.

Finally, we argue that direct model comparison is challenging, as each model has its own unique advantages. Some prioritize accuracy, others emphasize ease of implementation, and some provide intuitive insights or versatility.
The DEM approach has significant advantages, especially in the last three perspectives.
In this regard, the present work highlights the strengths and limitations of each approach, offering valuable guidance for selecting appropriate modeling techniques in granular flow simulations.

\section*{ACKNOWLEDGMENTS}

This work was supported by 2025 Hongik University Research Fund.

\section*{AUTHOR DECLARATIONS}
\subsection*{Conflict of Interest}
The authors have no conflicts to disclose.

\section*{DATA AVAILABILITY}
The data that support the findings of this study are available within the article.

\appendix
\section{Appendix}

\subsection{Derivation of the form of explicit Cauchy stress from an elastic-perfectly plastic model}
\label{sec_app:Cauchy_Stress}

In this section, we provide details on how to derive Eq. \eref{eqn:cauchy_stress_results}.
The procedures mostly involve tedious but direct mathematical manipulation, while a few key steps are summarized below.

First, substituting Eqs. \eref{eqn:generalized_hooke_law} and \eref{eqn:plastic_flow_rule}
to Eq. \eref{eqn:total_strain_plasticity}, 
\begin{equation}
    {{\dot{\epsilon }}^{\alpha \beta }}=\frac{{{{\dot{s}}}^{\alpha \beta }}}{2G}+\frac{1-2\nu }{3E}{{\dot{\sigma }}^{\gamma \gamma }}{{\delta }^{\alpha \beta }}+\dot{\lambda }\frac{\partial g}{\partial {{\sigma }^{\alpha \beta }}}. 
    \label{app:eqn_intermediate_0}
\end{equation}
Replacing $\dot{s}^{\alpha\beta}$ with $\dot\sigma^{\alpha \beta}-(1/3) \dot\sigma^{\gamma\gamma}\delta^{\alpha\beta}$, 
rearranging the equation in terms of $\dot\sigma^{\alpha \beta}$
, and introducing $K$ and $G$, we have 
\begin{equation}
    {{\dot{\sigma }}^{\alpha \beta }}=2G{{\dot{\varepsilon }}^{\alpha \beta }}+\left( \frac{3K-2G}{9K} \right){{\dot{\sigma }}^{\gamma \gamma }}{{\delta }^{\alpha \beta }}-2G\dot{\lambda }\frac{\partial g}{\partial {{\sigma }^{\alpha \beta }}},  
    \label{app:eqn_intermediate_1}
\end{equation}
where $K$ and $G$ are the constants defined in Eqs.\eref{eqn:plasticity_K} and \eref{eqn:plasticity_G}, respectively.

Next, we will eliminate $\sigma^{\gamma\gamma}$ in Eq. \eref{app:eqn_intermediate_1}. To this end, 
we multiply Eq. \eref{app:eqn_intermediate_0} by $\delta^{\alpha\beta}$
\begin{equation*}
    \left( {{{\dot{\epsilon }}}^{\alpha \beta }} \right) {{\delta }^{\alpha \beta }}=\left( \frac{{{{\dot{s}}}^{\alpha \beta }}}{2G}+\frac{1-2\nu }{3E}{{{\dot{\sigma }}}^{\gamma \gamma }}{{\delta }^{\alpha \beta }}+\dot{\lambda }\frac{\partial g}{\partial {{\sigma }^{\alpha \beta }}} \right){{\delta }^{\alpha \beta }},
\end{equation*}
which reduces to 
\begin{equation}
    {{\dot{\epsilon }}^{\gamma \gamma }}=\frac{1-2\nu }{3E}{{\dot{\sigma }}^{\gamma \gamma }}(3)+\dot{\lambda }\frac{\partial g}{\partial {{\sigma }^{\alpha \beta }}}{{\delta }^{\alpha \beta }}=\frac{1}{3K}{{\dot{\sigma }}^{\gamma \gamma }}+\dot{\lambda }\frac{\partial g}{\partial {{\sigma }^{\alpha \beta }}}{{\delta }^{\alpha \beta }}
    \label{app:eqn_intermediate_2}
\end{equation}
using the fact that the diagonal sum of the deviatoric stress is zero, i.e. $\dot{s}^{\gamma\gamma}=0$. 
Rearranging Eq. \eref{app:eqn_intermediate_2} in terms of $\dot\sigma^{\gamma\gamma}$, and substituting the result back into Eq. \eref{app:eqn_intermediate_1}, 
one arrives at
\begin{equation}
    {{\dot{\sigma }}^{\alpha \beta }}=2G{{\dot{\epsilon }}^{\alpha \beta }}+\left( \frac{3K-2G}{9K} \right)3K \left[{{\dot{\epsilon }}^{\gamma \gamma }}-\dot{\lambda }\frac{\partial g}{\partial {{\sigma }^{mn}}}{{\delta }^{mn}}\right]{{\delta }^{\alpha \beta }}-2G\dot{\lambda }\frac{\partial g}{\partial {{\sigma }^{\alpha \beta }}}
    \label{app:eqn_intermediate_3}
\end{equation}
Now, the remaining steps to derive Eq.~\eref{eqn:cauchy_stress_results} from Eq. \eref{app:eqn_intermediate_3} are straightforward.   

\subsection{Stabilization of SPH}
\label{sec_app:SPH_stabilizaton}

The SPH version of the momentum conservation in Eq.~\eref{eqn:SPH_vupdate_discretized} is often stabilized to avoid numerical oscillation and tensile instabilities. 
In this section, we brief the associated numerical techniques to address such issues.  
The stabilized version of the discretized momentum conservation in Eq.~\eref{eqn:SPH_vupdate_discretized} can be written as 
\begin{equation}
    \frac{Dv^{\alpha}_i}{Dt} = 
    \sum\limits_{j = 1}^N m_j 
    \left(
\frac{\sigma^{\alpha\beta}_i}{\rho_i^2}
+
\frac{\sigma^{\alpha\beta}_j}{\rho_i^2}-
\boxed{
\Pi_{ij}\delta^{\alpha\beta}
} + \boxed{f^\mathsf{n}_{ij}(R^{\alpha\beta}_i + R^{\alpha\beta}_j)}
    \right)
   \frac{\partial W_{ij}}{\partial x^{\beta}}+ b^{\alpha}
\label{eqn:SPH_vupdate_discretized_stabilized}
\end{equation}
where the boxed terms in Eq.~\eref{eqn:SPH_vupdate_discretized} highlight the stabilizing terms.  
The first boxed term is the artificial viscosity \cite{Monaghan:1992}, 
which is written as 
\begin{equation*}
    \Pi_{ij}=\left\{
    \begin{aligned}
&\frac{-\alpha_{\Pi}c\phi_{ij}+\beta_{\Pi}\phi^2}{0.5(\rho_i+\rho_j)},
\quad \mathrm{if} \;  (\mathbf{v}_i-\mathbf{v}_j)\cdot(\mathbf{x}_i-\mathbf{x}_j) < 0
\\
    &0,
    \quad \mathrm{otherwise.}
    \end{aligned}
    \right.
\end{equation*}
where 
\begin{equation*}
    \phi_{ij}= \frac{h(\mathbf{v}_i-\mathbf{v}_j)\cdot(\mathbf{x}_i-\mathbf{x}_j)}{|\mathbf{x}_i-\mathbf{x}_j|^2+0.01h^2}
\end{equation*}
and $c$ is the maximum sound speed in soil and calculated by $c=\sqrt{E/\rho}$ in the present study.  
The values of $\alpha_{\Pi}$ and $\beta_{\Pi}$ should be chosen according to particular applications. Here, $\alpha_{\Pi}$ and $\beta_{\Pi}$ are set to 0.1 and 0, respectively.

The second boxed term is a small repulsive force
\begin{equation*}
    f_{ij}= \frac{W_{ij}}{W_c}, 
\end{equation*} 
where $W_c$ is a constant that depends on initial particle spacing $\Delta d$ and smoothing length $h$. 

The repulsive force mimics inter-atomic repulsion and prevents neighboring particles from getting closer when in a state of tensile stress~\cite{Monaghan:2000}. 
For example, this artificial repulsive term only takes effect when particles clump, $r_{ij}< \Delta d$. 
In our study, $\Delta d$ varies depending on the particle resolution of the problem and the exponent $\mathsf{n}= 2.55$. 
The components of the artificial stress tensor $R^{\alpha \beta}_i$ for particle $i$ in the reference coordinate system $(x,y)$ is evaluated in the following way: 
\begin{enumerate}
    \item Take components of stress tensor of particle $i$ in the reference frame: $\sigma^{xx}_i,  \sigma^{yy}_i, \sigma^{xy}_i$.
    \item Calculate the diagonal $\sigma^{p,xx}_i, \sigma^{p,yy}_i$ components of stress tensor in the principal coordinate: 
    \begin{equation*}
        \sigma^{p,xx}_i = 
        \sigma^{xx}_i\cos^2\theta_i  + 2 \sigma^{xy}_i \cos\theta_i \sin\theta_i  + \sigma^{yy}_i \sin^2\theta_i 
    \end{equation*}
    \begin{equation*}
        \sigma^{p,yy}_i = 
        \sigma^{xx}_i\sin^2\theta_i  - 2 \sigma^{xy}_i \cos\theta_i \sin\theta_i  + \sigma^{yy}_i \cos^2\theta_i 
    \end{equation*}
    where the angle $\theta_i$ is defined by
    \begin{equation*}
        \tan 2\theta_i = \frac{2 \sigma^{xy}_i}{\sigma^{xx}_i - \sigma^{yy}_i}. 
    \end{equation*}
    \item Evaluate the diagonal components of the artificial stress tensor

    \begin{equation*}
    R^{p,xx}_{i}=\left\{
    \begin{aligned}
&-\epsilon \frac{\sigma^{p,xx}_i}{\rho_i^2},
\quad \mathrm{if} \; \sigma^{p,xx}_i>0
\\
    &0,
    \quad \mathrm{otherwise.}
    \end{aligned}
    \right.
\end{equation*}

\item Finally, compute the each component $R^{\alpha \beta}_i$ as
\begin{equation*}
    R^{xx}_i = R^{p,xx}_i \cos^2 \theta_i + R^{p,yy} \sin^2 \theta_i 
\end{equation*}
\begin{equation*}
    R^{yy}_i = R^{p,xx}_i \sin^2 \theta_i + R^{p,xx} \cos^2 \theta_i 
\end{equation*}
\begin{equation*}
    R^{xy}_i = (R^{p,xx}_i-R^{p,yy}_i) \sin \theta_i \cos \theta_i 
\end{equation*}

\end{enumerate}

It is known that above artiﬁcial stress terms introduce only negligible errors into the calculation results~\cite{Monaghan:2000,Monaghan:2001}.

\printbibliography

@article{Monaghan_2005_Review,
    author = {J. J. Monaghan},
    title = {Smoothed particle hydrodynamics},
    journal = {Recent Progress in Physics},
    year = {2005}, 
    pages={1703},
    vol={68},
    doi={https://doi.org/10.1088/0034-4885/68/8/R01}
}

@article{Jkim2021,
    author = {J. Kim and J. D. Park},
    title = {A thixotropic fluid flow around two sequentially aligned spheres},
    journal = {orean Journal of Chemical Engineering},
    year = {2021}, 
    pages={1460-1468}, 
    vol={38},
    doi={https://doi.org/10.1007/s11814-021-0780-x}
}

@article{DEMtimeStpper,
  title = {Computer "Experiments" on Classical Fluids. I. Thermodynamical Properties of Lennard-Jones Molecules},
  author = {Loup Verlet},
  journal = {Physics Review Journals},
  volume = {159},
  pages = {98-103},
  year = {1967},
  publisher = {American Physical Society},
  doi = {https://doi.org/10.1103/PhysRev.159.98},
}

@article{RotatingDrum0,
  title = {Granular flow regimes in rotating drums from depth-integrated theory},
  author = {C.-Y. Hung and C. P. Stark and H. Capart},
  journal = {Physical Review E},
  volume = {93},
  issue = {3},
  pages = {030902},
  year = {2016},
  publisher = {American Physical Society},
  doi = {https://doi.org/10.1103/PhysRevE.93.030902},
}

@article{EPSD,
title = {Assessment of rolling resistance models in discrete element simulations},
journal = {Powder Technology},
volume = {206},
number = {3},
pages = {269-282},
year = {2011},
issn = {0032-5910},
doi = {https://doi.org/10.1016/j.powtec.2010.09.030},
author = {J. Ai and J.-F. Chen and J. M. Rotter and J. Y. Ooi}
}

@article{ANGUS2020290,
title = {Calibrating friction coefficients in discrete element method simulations with shear-cell experiments},
journal = {Powder Technology},
volume = {372},
pages = {290-304},
year = {2020},
issn = {0032-5910},
doi = {https://doi.org/10.1016/j.powtec.2020.05.079},
author = {A. Angus and L. A. A. Yahia and R. Maione and M. Khala and C. Hare and A. Ozel and R. Ocone},
}

@article{Juarez_2011,
doi = {https://doi.org/10.1088/1367-2630/13/5/053055},
year = {2011},
month = {may},
volume = {13},
number = {5},
pages = {053055},
author = {G. Juarez and P. Chen and R M. Lueptow},
title = {Transition to centrifuging granular flow in rotating tumblers: a modified Froude number},
journal = {New Journal of Physics},
}

@article{POURANDI2024120176,
title = {A mathematical model for the dynamic angle of repose of a granular material in the rotating drum},
journal = {Powder Technology},
volume = {446},
pages = {120176},
year = {2024},
issn = {0032-5910},
doi = {https://doi.org/10.1016/j.powtec.2024.120176},
author = {S. Pourandi and P. C. {van der Sande} and T. Weinhart and I. Ostanin}
}

@book{NumericalMethod,
  title     = "Numerical Methods for Ordinary Differential Equations",
  author    = "J. C. Butcher",
  year      = 2003,
  publisher = "John Wiley \& Sons",
  address   = "London"
}

@article{Chau:1999,
    author = {Chau, K.T. and Wong, R.H.C. and Liu, J. and Wu, J.J. and Lee, C.F.},
    title = {Shape Effects On the Coefficient of Restitution During Rockfall Impacts},
    volume = {All Days},
    series = {ISRM Congress},
    pages = {ISRM-9CONGRESS-1999-111},
    year = {1999},
    month = {08},
}

@article{HLOSTA2018222,
title = {Experimental determination of particle–particle restitution coefficient via double pendulum method},
journal = {Chemical Engineering Research and Design},
volume = {135},
pages = {222-233},
year = {2018},
issn = {0263-8762},
doi = {https://doi.org/10.1016/j.cherd.2018.05.016},
url = {https://www.sciencedirect.com/science/article/pii/S0263876218302478},
author = {J. Hlosta and D. Žurovec and J. Rozbroj and Á. Ramírez-Gómez and J. Nečas and J. Zegzulka},
}

@article{JKim:2018,
	author = {J. B. Freund and J. Kim and R. H. Ewoldt},
	journal = {Journal of Non-Newtonian Fluid Mechanics},
	doi = {https://doi.org/10.1016/j.jnnfm.2018.03.013},
	pages = {71--82},
	title = {Field sensitivity of flow predictions to rheological parameters},
	volume = {257},
	year = {2018}}

@article{Renardy:2021,
	author = {M. Renardy and B. Thomases},
	journal = {Journal of Non-Newtonian Fluid Mechanics},
	doi = {https://doi.org/10.1016/j.jnnfm.2021.104573},
	pages = {104573},
	title = {A mathematician's perspective on the {O}ldroyd {B} model: Progress and future challenges},
	volume = {293},
	year = {2021}}

@ARTICLE{Park:2024,
  author = {H.-J. Park and J. Kim and H.-J. Kim},
  journal={Physics of Fluids}, 
  title={Segment-based wall treatment model for heat transfer rate in smoothed particle hydrodynamics}, 
  year={2024},
  volume={36},
  number={7},
  pages={077106 },
  doi={https://doi.org/10.1063/5.0211482}}

@ARTICLE{Barr,
  author={Barr},
  journal={IEEE Computer Graphics and Applications}, 
  title={Superquadrics and Angle-Preserving Transformations}, 
  year={1981},
  volume={1},
  number={1},
  pages={11-23},
  doi={https://doi.org/10.1109/MCG.1981.1673799}}

@article{DEM_Sullivan,
	author = {C. O. Sullivan},
	journal = {International Journal of Geomechanics},
	doi = {https://doi.org/10.1061/(ASCE)GM.1943-5622.000002},
	pages = {449-464},
	title = {Particle-Based Discrete Element Modeling: Geomechanics Perspective},
	volume = {11},
    issue = {6},
	year = {2011}}

@article{DEM_Cleary,
	author = {P. W. Cleary and N. Stokes and J. Hurley},
	journal = {Computational Techniques and Applications},
	pages = {1-7},
	title = {Efﬁcient collision detection
for three dimensional super-ellipsoidal particles},
	year = {1997}}

@inbook{DEM_Jonhson,
	author = {K.L Jonhson},
	booktitle = {In: Contact Mechanics.},
	doi = {https://doi.org/10.1017/CBO9781139171731.005},
	pages = {84-106},
	title = {Normal contact of elastic solids – Hertz theory},
	year = {1985},
    publisher={Cambridge University Press}}

@article{Cundall,
	author = {P. A. Cundall and O. D. L. Strack},
	journal = {Geotechnique},
	doi = {https://doi.org/10.1680/geot.1979.29.1.47},
	pages = {47-65},
	title = {A discrete numerical model for granular assemblies},
	volume = {29},
        number = {1},
	year = {1979}}

@article{Liggght,
	author = {C. Kloss and C. Goniva and Allice Konig and Stefan and Amberger and S. Pirker},
	journal = {Progress in Computational Fluid Dynamics, an International Journal},
	doi = {https://doi.org/10.1504/PCFD.2012.047457},
	pages = {140-152},
	title = {Models, algorithms and validation for opensource {DEM} and {CFD-DEM}},
	volume = {12},
    number = {2-3},
	year = {2012}}

@article{DEM_superquadratic,
	author = {A. Podlozhnyuk and S. Pirker and C. Kloss},
	journal = {Computational Particle Mechanics},
	doi = {https://doi.org/10.1007/s40571-016-0131-6
Efficient},
	pages = {101–118},
	title = {Efficient implementation of superquadric particles in Discrete Element Method within an open-source framework},
	volume = {4},
	year = {2017}}

@article{SPH_Bui:01,
	author = {H. H. Bui and R. Fukagawa and K. Sako and S. Ohno},
	journal = {Numerical Analysis and Methodd in Geomechanics},
	doi = {https://doi.org/10.1002/nag.688},
	pages = {1537-1570},
	title = {Lagrangian meshfree particles method (SPH) for large deformation and failure flows of geomaterial using elastic–plastic soil constitutive model},
	volume = {32},
	year = {2008}}

@article{JKim:2019,
	author = {J. Kim and P. K. Singh and J. B. Freund and R. H. Ewoldt},
	journal = {Journal of Non-Newtonian Fluid Mechanics},
	doi = {https://doi.org/10.1016/j.jnnfm.2019.07.002},
	pages = {104138},
	title = {Uncertainty propagation in simulation predictions of generalized Newtonian fluid flows},
	volume = {271},
	year = {2019}}

@article{JKim:2023,
	author = {J. Kim},
	journal = {Applied Mathematical Modelling},
	doi = {https://doi.org/10.1016/j.apm.2022.10.044},
	pages = {453-469},
	title = {Adjoint-based sensitivity analysis of viscoelastic fluids at a low deborah number},
	volume = {115},
	year = {2023}}

@article{kim2023application,
  title={Application of artificial neural networks using sequential prediction approach in indoor airflow prediction},
  author={Kim, MinHo and Park, Hyung-Jun},
  journal={Journal of Building Engineering},
  volume={69},
  pages={106319},
  year={2023},
  publisher={Elsevier}
}

@article{kim2023direct,
  title={Direct imposition of the wall boundary condition for weakly compressible flows in three-dimensional smoothed particle hydrodynamics simulations},
  author={Kim, Imgyu and Park, Hyung-Jun},
  journal={Physics of Fluids},
  volume={35},
  number={11},
  year={2023},
  publisher={AIP Publishing}
}

@article{Park:2023,
  title = {A New {{SPH-FEM}} Coupling Method for Fluid\textendash Structure Interaction Using Segment-Based Interface Treatment},
  author = {H.-J. Park and H.-D. Seo},
  year = {2023},
  journal = {Engineering with Computers},
  issn = {1435-5663},
  doi ={https://doi.org/10.1007/s00366-023-01856-1}
}

@article{huIncompressibleMultiphaseSPH2007,
  title = {An Incompressible Multi-Phase {{SPH}} Method},
  author = {X. Y. Hu and N. A. Adams},
  year = {2007},
  journal = {Journal of computational physics},
  volume = {227},
  number = {1},
  pages = {264--278},
  publisher = {{Elsevier}},
  doi = {https://doi.org/10.1016/j.jcp.2007.07.013},
}

@article{Monaghan:1992,
  title = {Smoothed particle hydrodynamics},
  author = {J. J. Monaghan},
  year = {1992},
  journal = {Annual Review of Astronomy and Astrophysics},
  volume = {30},
  pages = {543–574},
  doi = {https://doi.org/10.1146/annurev.aa.30.090192.002551} 
}

@article{Monaghan:2000,
  title = { {SPH} without a tensile instability},
  author = {J. J. Monaghan},
  year = {2000},
  journal = {Journal of Computational Physics},
  volume = {159},
  pages = {290-311},
  doi = {https://doi.org/10.1006/jcph.2000.6439} 
}

@article{Monaghan:2001,
  title = {{SPH} elastic dynamics},
  author = {J. P. Gray and J. J. Monaghan and R. P. Swift},
  year = {2001},
  journal = {Computer Methods in Applied Mechanics and Engineering},
  volume = {190},
  pages = {6641-6662},
  doi = {https://doi.org/10.1016/S0045-7825(01)00254-7} 
}

@article{CollapseExperiment,
  title = {Failure Mechanism of True 2{D} Granular Flows},
  author = {C. T. Nguyen and H. H. Bui and R. Fukagawa},
  year = {2015},
  journal = {Joural of Chemical Engineering of Japan},
  volume = {48},
  pages = {395-402},
  doi = {https://doi.org/10.1252/jcej.14we358} 
}

@book{liu2003smoothed,
  title={Smoothed particle hydrodynamics: a meshfree particle method},
  author={Liu, Gui-Rong and Liu, Moubin B},
  year={2003},
  publisher={World scientific}
}

@article{bui2021smoothed,
  title={Smoothed particle hydrodynamics (SPH) and its applications in geomechanics: From solid fracture to granular behaviour and multiphase flows in porous media},
  author={Bui, Ha H and Nguyen, Giang D},
  journal={Computers and Geotechnics},
  volume={138},
  pages={104315},
  year={2021},
  publisher={Elsevier}
}

@article{feng2021large,
  title={Large deformation analysis of granular materials with stabilized and noise-free stress treatment in smoothed particle hydrodynamics (SPH)},
  author={Feng, Ruofeng and Fourtakas, Georgios and Rogers, Benedict D and Lombardi, Domenico},
  journal={Computers and Geotechnics},
  volume={138},
  pages={104356},
  year={2021},
  publisher={Elsevier}
}

@article{nguyen2017new,
  title={A new SPH-based approach to simulation of granular flows using viscous damping and stress regularisation},
  author={Nguyen, Cuong T and Nguyen, Chi T and Bui, Ha H and Nguyen, Giang D and Fukagawa, Ryoichi},
  journal={Landslides},
  volume={14},
  pages={69--81},
  year={2017},
  publisher={Springer}
}

\end{document}